\documentclass[a4paper,11pt]{article}
\usepackage{jmehppub}

\usepackage[T1]{fontenc}
\usepackage[utf8]{inputenc}
\usepackage{mathtools}
\usepackage{physics}
\usepackage{dsfont,tensor,mathbbol,mathdots}
\usepackage{graphicx,subcaption}
\usepackage[export]{adjustbox}
\usepackage[dvipsnames,table]{xcolor}
\usepackage[normalem]{ulem}
\usepackage[centertableaux]{ytableau}
\usepackage{sidenotes,comment,booktabs,placeins}
\usepackage{tikz,tikz-3dplot}
\usetikzlibrary{calc,arrows.meta,decorations.markings,decorations.pathmorphing,decorations.pathreplacing,patterns}
\usepackage{setspace}
\colorlet{darkgreen}{green!50!black}

\newcommand{\be}{\begin{equation}}
\newcommand{\ee}{\end{equation}}
\newcommand{\beq}{\begin{equation}}
\newcommand{\eeq}{\end{equation}}

\title{The Shape of the Strongly Coupled MQM Ground State}
\author[a]{Alexander Frenkel}
\affiliation[a]{Simons Center for Geometry and Physics, Stony Brook University, Stony Brook, NY 11794, USA}
\emailAdd{afrenkel@scgp.stonybrook.edu}

\abstract{We study a surprisingly efficient family of variational ans\"atze for the ground state of bosonic matrix quantum mechanics at large $N$. These are ans\"atze whose norm-squared probability distributions are themselves matrix integrals with single-trace actions, and so have their own independent `t Hooft expansions. They are efficient in the sense that optimizing over a small number of parameters produces many digits of accuracy for the energy, single-trace moments, and connected correlators. For infinite-$N$ planar data our results rapidly converge to within published bootstrap intervals and stabilize to a few digits of precision beyond them. We also demonstrate that we can efficiently capture $1/N$ corrections by computing two-point functions. These are matched against resummed quantum perturbation theory and Monte Carlo simulations.}
\date{}

\begin{document}
\maketitle

\section{Introduction}

Large-$N$ Matrix Quantum Mechanics (MQM) at strong `t Hooft coupling is a promising playground for understanding holography and non-perturbative theories of quantum gravity~\cite{tHooft:1974Planar,Ginsparg:1993Lectures,Das:1990StringField,Jevicki:1980Collective,Gross:1990Circle,Gross:1991Fermionic,Banks:1997BFSS,Itzhaki:1998Supergravity,Berenstein:2002BMN,Lin:2025TASI}\footnote{See \cite{Lin:2025TASI} in particular for a very modern review.}. It is therefore interesting to solve the theory to the extent possible, and in particular understand the structure of the ground state. Because multi-matrix quantum mechanics is a strongly coupled system of infinitely many degrees of freedom, this problem is hard. Recent works~\cite{Michel:2026GroundState,Bajaj:2026SU2,Sanjurjo:2026BoundStates} have used modern computational tools to numerically solve $N = 2$ BFSS,\footnote{The existence of the $SU(2)$ threshold bound state was established by Sethi and Stern~\cite{Sethi:1998BoundStates}.} and even this progress is enough to understand the emergence of graviton scattering~\cite{Herderschee:2024ThreePoint}.

In this paper we attack the (bosonic) large-$N$ problem directly, searching for explicit wavefunctions $\Psi(X^i)$ of $D$ bosonic matrices $X^i$ that capture the structure of the ground state via the variational approach. Unlike~\cite{Han:2020Bootstrap,Lin:2026Precision,deMelloKoch:2022Optimization,Mathaba:2024MasterField,Rodrigues:2025Glueballs,Liu:2026CollectiveBootstrap} the goal is to get at a compactly-expressed wavefunction in the original matrix variables, and not only expectation values of observables\footnote{This distinction is somewhat counter to the point of view of~\cite{deMelloKoch:2022Optimization,Jevicki:1983LoopSpace}, which is that loop spaces \textit{are} the correct Hilbert space and wavefunction to consider.}, although we find as an unexpected benefit that energies and observables converge remarkably well. First, having the explicit wavefunction, especially if a good approximation is compactly stated (as we find here), makes recovery of many moments and expectation values extremely computationally cheap. Second, once we have an explicit wavefunction we can begin to study quantum information theoretic properties (such as entanglement entropy~\cite{Hartnoll:2015Entanglement,Han:2020DeepQuantumGeometry,Hampapura:2021TargetSpace,Das:2020BulkEntanglement,Das:2021GaugeInvariantEntanglement,Das:2022Finiteness,Frenkel:2022QuantumHallEntanglement,Frenkel:2023AreaLaws,Fliss:2025MinimalAreas,Fliss:2026EntanglementReview}) of MQM ground states directly in the regime of strong `t Hooft coupling, which has previously been mostly inaccessible.

The hypothesis is simple and mostly inspired by the philosophy elucidated in~\cite{Fliss:2026Topology} -- if the `t Hooft expansion should be taken seriously as the non-perturbative definition of the worldsheet (a point of view bolstered by explicit constructions in special settings in~\cite{Aharony:2006Worldsheet,Gopakumar:2022Triality,Gopakumar:2026FeynmanStrings,Gaiotto:2025Categorical}), then wavefunctions that generate their own `t Hooft expansions are interesting objects themselves. We therefore consider wavefunctions that have the following schematic form:
\begin{equation}\label{eqn:intro-single-trace}
\Psi(X^i) = \mathcal{N}\exp(-g_2 \sum_i \Tr[(X^i)^2] - g_{4;ijkl}\Tr[X^iX^jX^kX^l] + \ldots),
\end{equation}
With  the approximation made finite by truncating in short words (we call the length of the longest word $L$). This idea is not new~\cite{Schwartz:1999Computations,Agarwal:2003Fisher}, and in fact in 1999 Schwartz\footnote{In a paper that has apparently lain unfairly undisturbed since then. The author would not have found it in a literature search without modern AI tools.}~\cite{Schwartz:1999Computations} already noted the remarkably fast numerical convergence for large-$N$ planar moments as the cumulant cutoff is increased. Karabali, Kim, and Nair have applied an idea similar in spirit to great effect in 2+1d Yang Mills~\cite{Karabali:1998Vacuum}.

The prior work of \cite{deMelloKoch:2022Optimization} is essential to us -- to optimize ans\"atze like \eqref{eqn:intro-single-trace} we use an adaptation of their algorithm. We write the planar energy as a function of the moments of our trial wavefunction, which is the starting point of \cite{deMelloKoch:2022Optimization}. In some sense, when viewed as generating functions of moments, our single-trace ans\"atze are simply an extremely restricted search through their space of moments (they optimize over thousands of parameters, we optimize over a couple of dozen). Part of the surprising result is that this restricted search seems to not cost any accuracy, and in fact earns us several digits of precision by allowing us to work with a much higher effective cap on moments.

In this paper we generalize this ansatz in a way that captures $1/N$ corrections (especially connected correlation functions, although in this paper we restrict our attention to connected two-point functions) by integrating in auxiliary matrices $A$:
\begin{equation}\label{eqn:intro-auxiliary}
\Psi(X^i) = \mathcal{N} \exp(-\Tr[f(X^i)])\int d^{N^2}A \exp(-\Tr[h(X^i,A)]) \ldots,
\end{equation}
where the ellipsis suppresses potentially arbitrarily many (bosonic or Grassmann) auxiliary variables. This generalization of the variational ansatz appears to be new. Integrating out $A$ in \eqref{eqn:intro-auxiliary} would produce an infinite tower of multi-trace corrections, but in \S\ref{sec:connected-covariances} we show that optimizing a few auxiliary parameters substantially improves the covariance predictions. However, we also note that one of the surprises of this work is that even the planar single-trace state \eqref{eqn:intro-single-trace}, having optimized only the $O(N^2)$ energy, captures covariances extremely well: for example, for the two-matrix model at moderate mass $\mu=1$, the median and largest relative discrepancies across all ten tested covariances are $2.41\times10^{-3}$ and $3.32\times10^{-3}$, respectively, compared with resummed QPT estimates (see \S\ref{sec:two-matrix-covariances}). A couple of auxiliary couplings in the structure of \eqref{eqn:intro-auxiliary} then improve even on \textit{this} agreement by more than an order of magnitude.

That wavefunctions like \eqref{eqn:intro-single-trace}, \eqref{eqn:intro-auxiliary}, and their generalizations capture the structure of the strongly-coupled MQM ground state is the main idea of this paper. Everything below is simply providing evidence for this idea by numerically minimizing these ans\"atze and demonstrating their rapid convergence. We lay out precisely our variational search space and optimization strategy in \S\ref{sec:ansatze} (optimization algorithm details are in Appendix \ref{app:optimization}). For planar data we match against MQM bootstrap~\cite{Han:2020Bootstrap,Lin:2023D0Bootstrap,Lin:2026Precision}, especially the precision bootstrap work of Lin and Zheng~\cite{Lin:2026Precision}. These rigorous bounds are invaluable for independently verifying the effectiveness of these trial state families. For covariances precision bootstrap does not exist, so we resort to resummed quantum perturbation theory (QPT) (especially the work of Kabat and Lifschytz~\cite{Kabat:2000Approximations}) and Monte Carlo simulations to independently check our predictions for the connected correlators (details are in Appendices \ref{app:resummation} and \ref{app:monte-carlo}).

\section{Single-Trace Ans\"atze and Variational Setup}\label{sec:ansatze}

The main idea is to consider variational wavefunctions whose squared norms define probability distributions with a good 't Hooft limit. For a collection of $D$ Hermitian matrices $X^i$, $i=1,\ldots,D$, normalized to have $O(1)$ eigenvalues, the basic ansatz is
\begin{equation}\label{eqn:single-trace}
\Psi_{\boldsymbol g}(X)=\mathcal N_{\boldsymbol g}e^{-\Tr f_{\boldsymbol g}(X)},
\qquad f_{\boldsymbol g}(X)=N\sum_wg_w\,w(X).
\end{equation}
The sum runs over nonempty words in the $X^i$, with cyclic duplicates removed; we vary a chosen finite set of coefficients $\boldsymbol g=(g_w)$ and set all others to zero. We allow complex coefficients, requiring only that $\int dX\,e^{-2\operatorname{Re}\Tr f_{\boldsymbol g}(X)}<\infty$. The factor of $N$ gives the usual planar scaling at fixed $\boldsymbol g$, and $\mathcal N_{\boldsymbol g}$ normalizes the state.

We can effectively include multi-trace terms by integrating over an auxiliary Hermitian matrix $A$, with the same eigenvalue normalization:
\begin{equation}\label{eqn:basic-auxiliary-ansatz}
\begin{aligned}
\Psi_{\boldsymbol g,\boldsymbol\eta}(X)
&=\mathcal N_{\boldsymbol g,\boldsymbol\eta}e^{-\Tr f_{\boldsymbol g}(X)}
\int d^{N^2}A\,e^{-\Tr h_{\boldsymbol\eta}(X,A)},\\
h_{\boldsymbol\eta}(X,A)&=N\sum_{w\in\mathcal V}\eta_w\,w(X,A).
\end{aligned}
\end{equation}
Here $\mathcal V$ is a chosen finite set of cyclically inequivalent words containing at least one $A$; words involving only the $X^i$ are already included in $f_{\boldsymbol g}$. The coefficients $\boldsymbol\eta=(\eta_w)$ provide additional complex variational parameters, subject to normalizability of the resulting wavefunction. More auxiliary matrices can be introduced in the same way, but we will find that a few short words in the exponents already work remarkably well. We also allow Grassmann (anticommuting) auxiliary matrices, as in \S\ref{sec:one-matrix-covariances}; these integration variables do not add physical fermions to the MQM.

\subsection{Optimization via Schwinger-Dyson Cumulant Truncation}

Collective field theory formulates the large-$N$ ground-state problem as a constrained minimization in `loop space,' the space of single-trace moments \cite{Jevicki:1980Collective,Jevicki:1984LoopSpaceII,deMelloKoch:2022Optimization,Liu:2026CollectiveBootstrap}. Our ans\"atze \eqref{eqn:single-trace} and \eqref{eqn:basic-auxiliary-ansatz} generate an infinite hierarchy of moments from a small number of parameters, the coefficients of words in the exponents. They therefore provide a highly restricted variational search in loop space. For each choice of coefficients, we solve a truncation of the Schwinger--Dyson (SD) equations for the associated matrix integrals, organized in free cumulants. Note, crucially, that the SD equations are \textit{not} those obtained from the MQM path integral -- they are those obtained from the matrix integral obtained by taking the norm squared of our ansatz $\Psi$. We use the resulting approximate moments to evaluate the energy expectation value and minimize it over the wavefunction coefficients.

Free probability organizes the planar single-trace moments of matrix integrals and MQM as a noncommutative probability distribution \cite{Akant:2002Entropy,Agarwal:2003Fisher}. Normalizing the matrices to have $O(1)$ eigenvalues, we define the planar moments by
\begin{equation}\label{eqn:planar-moments}
m_{i_1\cdots i_n}\equiv\lim_{N\to\infty}\frac1N\left\langle\Tr\!\left(X^{i_1}\cdots X^{i_n}\right)\right\rangle_\Psi.
\end{equation}
For a word, we also use the abbreviated moment notation and denote its free cumulant by
\begin{equation}\label{eqn:word-cumulant}
\begin{gathered}
w=X^{i_1}\cdots X^{i_n},\\
m(w)=m_w=m_{i_1\cdots i_n},\qquad
\kappa_w\equiv\kappa_{i_1\cdots i_n}.
\end{gathered}
\end{equation}
We extend $m$ linearly to matrix polynomials. The moment--cumulant relation below defines these quantities recursively in word length \cite{Speicher:1994Cumulants}: the partition containing a single block contributes $\kappa_w$, while every other partition contributes products of cumulants associated with shorter words.
\begin{equation}\label{eqn:free-moment-cumulant}
m_{i_1\cdots i_n}=\sum_{\pi\in NC(n)}\prod_{\{b_1<\cdots<b_r\}\in\pi}\kappa_{i_{b_1}\cdots i_{b_r}},
\end{equation}
where $NC(n)$ denotes the noncrossing partitions of $\{1,\ldots,n\}$. These cumulants are the `planar connected parts' of Halpern and Schwartz \cite{Halpern:1999FreeAlgebra}; they describe correlations within a single trace and are distinct from connected correlators of several traces. We write $m_{\boldsymbol\kappa}(w)$ for the moment of a word $w$ reconstructed from the free cumulants using \eqref{eqn:free-moment-cumulant}, and extend this notation linearly to polynomials.

We introduce a word length cutoff $L$ by retaining only words with $|w|\leq L$ in \eqref{eqn:single-trace}. We consider a bosonic Hamiltonian
\begin{equation}\label{eqn:bosonic-hamiltonian}
H=\frac{1}{2N}\sum_{i=1}^D\Tr\Pi_i^2+N\Tr V(X^i),
\end{equation}
with conjugate momenta $\Pi_i$ and polynomial potential $V$. We define the energy per $N^2$ at finite $N$ and express its planar limit in terms of the moments of the trial state:
\begin{equation}\label{eqn:variational-energy}
\begin{aligned}
\mathcal E_N(\boldsymbol q)
&\equiv\frac{\langle\Psi_{\boldsymbol g}|H|\Psi_{\boldsymbol g}\rangle}{N^2\langle\Psi_{\boldsymbol g}|\Psi_{\boldsymbol g}\rangle},\\
\mathcal E(\boldsymbol\kappa,\boldsymbol q)
&=\lim_{N\to\infty}\mathcal E_N(\boldsymbol q)\\
&=m_{\boldsymbol\kappa}(V)+\frac12\sum_{i=1}^D\sum_{\substack{u,v\\1\leq |u|,|v|\leq L}}g_u^*g_v\,m_{\boldsymbol\kappa}\!\left((\mathcal D_i u)^\dagger\mathcal D_i v\right).
\end{aligned}
\end{equation}
Here $\boldsymbol q=(\operatorname{Re}\boldsymbol g,\operatorname{Im}\boldsymbol g)$. The matrix polynomial $\mathcal D_iw=\sum_{w=aX^ib}ba$ is obtained by differentiating $\Tr w$ with respect to $X^i$: for each occurrence of $X^i$, we write $w=aX^ib$, where $a$ and $b$ are the ordered subwords before and after that occurrence, respectively. Either may be empty, in which case it represents the identity matrix. The order $ba$ follows from cyclicity, $\Tr(a\,\delta X^i\,b)=\Tr(ba\,\delta X^i)$. For example, $\mathcal D_X(X^2Y)=XY+YX$. Thus both the kinetic and potential energies are functions of the moments; the kinetic contribution requires words of length at most $2L-2$.

de Mello Koch et al. \cite{deMelloKoch:2022Optimization} minimize the collective potential by parametrizing the moments with finite matrices, which enforce the loop-space positivity constraints, and differentiating with respect to their independent entries. We use their minimization strategy, except we must first embed our coupling coordinates $\boldsymbol q$ into their space of moments. We then apply gradient minimization to the exponent coefficients $\boldsymbol q$. Following Schwartz \cite{Schwartz:1999Computations}, we introduce a separate cumulant cutoff $K$ by setting $\kappa_w=0$ for $|w|>K$. At each trial $\boldsymbol q$, we determine the independent real components $\boldsymbol\kappa$ of the retained cumulants from a finite selection of the SD equations, denoted $\boldsymbol F_K(\boldsymbol\kappa,\boldsymbol q)=0$. For the real two- and nine-matrix trial states used below, we minimize the SD-rewritten energy $\mathcal E_{\rm SD}$ defined in \eqref{eqn:algorithm-energy}. It equals \eqref{eqn:variational-energy} when the required SD identities hold, but the selected equations at finite $K$ need not enforce that equality. We initialize the SD solve with Gaussian cumulants. As we vary $\boldsymbol q$, we use the solution for $\boldsymbol\kappa$ at the preceding coefficient values as the initial guess for the next solve. On a solution branch where $J=\partial_{\boldsymbol\kappa}\boldsymbol F_K$ is nonsingular, differentiating the SD equations gives the gradient of this objective function:
\begin{equation}\label{eqn:sd-energy-gradient}
\begin{gathered}
e_{L,K}(\boldsymbol q)=\mathcal E_{\rm SD}(\boldsymbol\kappa(\boldsymbol q),\boldsymbol q),\qquad
J^{\mathsf T}\boldsymbol z=\nabla_{\boldsymbol\kappa}\mathcal E_{\rm SD},\\
\nabla_{\boldsymbol q}e_{L,K}=\partial_{\boldsymbol q}\mathcal E_{\rm SD}-(\partial_{\boldsymbol q}\boldsymbol F_K)^{\mathsf T}\boldsymbol z.
\end{gathered}
\end{equation}
The partial derivative $\partial_{\boldsymbol q}\mathcal E_{\rm SD}$ holds $\boldsymbol\kappa$ fixed, while the auxiliary vector $\boldsymbol z$ is defined by this linear equation. We use this gradient to update the coefficients and repeat the SD solution. The cutoff $L$ specifies the variational family, while $K$ specifies the approximation used to evaluate its moments.

\subsection{Optimization of Ans\"atze with Auxiliaries}

We use auxiliaries to vary covariances while holding the optimized planar moments fixed. For an action quadratic in an ordinary Hermitian auxiliary $A$, the integral in \eqref{eqn:basic-auxiliary-ansatz} gives
\begin{equation}\label{eqn:gaussian-auxiliary}
\begin{aligned}
\Tr h_{\boldsymbol\eta}(X,A)&=\frac N2\Tr\!\left(A\,\mathcal M_{\boldsymbol\eta}(X)[A]\right),\\
\int d^{N^2}A\,e^{-\Tr h_{\boldsymbol\eta}(X,A)}&\propto\left[\det\mathcal M_{\boldsymbol\eta}(X)\right]^{-1/2}.
\end{aligned}
\end{equation}
The determinant is over the $N^2$ real components of $A$, and the identity requires a convergent Gaussian integral. The contribution $\frac12\log\det\mathcal M_{\boldsymbol\eta}$ to the exponent generates single- and double-trace terms. Expanding about the optimized planar moments, we absorb the constant into the normalization and compensate the linear terms by adjusting the single-trace coefficients. This leaves a quadratic correction to the trace fluctuations.

Choose a finite set of independent real trace observables $T_a=\Tr p_a(X)$, where $p_a$ are Hermitian linear combinations of words, and write $\delta T_a=T_a-Nm(p_a)$. At a regular stable planar saddle, large-$N$ counting gives
\begin{equation}\label{eqn:trace-fluctuation-scaling}
\left\langle\delta T_{a_1}\cdots\delta T_{a_r}\right\rangle_c
=O\!\left(N^{2-r}\right),\qquad r\geq2,
\end{equation}
where the subscript $c$ denotes the connected part. Connected correlators of more than two trace fluctuations therefore vanish at leading order, so their joint distribution is Gaussian and determined by its covariance matrix. The accuracy of the computed covariance still depends on the planar input and the set of trace observables retained in the optimization. The compensated wavefunction and its leading covariance matrix therefore take the form
\begin{equation}\label{eqn:auxiliary-covariance}
\Psi\propto\Psi_{\boldsymbol g}\exp\!\left[-\frac12\delta\boldsymbol T^{\mathsf T}B_{\boldsymbol\eta}\delta\boldsymbol T\right],
\qquad C_{\boldsymbol\eta}=\left(C_0^{-1}+2\operatorname{Re}B_{\boldsymbol\eta}\right)^{-1}.
\end{equation}
Here $B_{\boldsymbol\eta}$ is the symmetric matrix of quadratic coefficients generated by the determinant, and $C_0$ is the leading covariance matrix of the single-trace state. We compute $C_0$ by differentiating its SD equations with respect to single-trace sources, using the same Jacobian as in \eqref{eqn:sd-energy-gradient}.

The energy follows from the quadratic expansion of the collective Hamiltonian \cite{Jevicki:1984MasterVariables,deMelloKoch:2022Optimization}; this framework also yields order-one thermal trace correlators \cite{Jevicki:2025Thermofield}. After averaging unvaried fluctuations with their conditional distribution in the single-trace state, let $\Omega$ and $R$ denote its kinetic and potential matrices on the retained observables. In particular, $\Omega_{ab}=\sum_i m((\mathcal D_i p_a)(\mathcal D_i p_b))$; $R$ includes the contribution from the change of integration measure from matrices to traces. The optimized single-trace state can be taken to have constant phase: a position-dependent phase adds nonnegative kinetic energy without changing the potential energy. The change in the $O(N^0)$ energy is
\begin{equation}\label{eqn:auxiliary-energy}
\begin{aligned}
H_{\rm fluc}&=-\frac12\nabla_{\delta T}^{\mathsf T}\Omega\nabla_{\delta T}+\frac12\delta\boldsymbol T^{\mathsf T}R\delta\boldsymbol T,\\
\Delta E_1(\boldsymbol\eta)&=\frac18\operatorname{tr}\!\left[\Omega(C_{\boldsymbol\eta}^{-1}-C_0^{-1})\right]+\frac12\operatorname{tr}\!\left[R(C_{\boldsymbol\eta}-C_0)\right]\\
&\quad+\frac12\operatorname{tr}\!\left[(\operatorname{Im}B_{\boldsymbol\eta})\Omega(\operatorname{Im}B_{\boldsymbol\eta})C_{\boldsymbol\eta}\right].
\end{aligned}
\end{equation}
Here the lowercase $\operatorname{tr}$ (not to be confused with the uppercase $\Tr$ trace over color space) is the ordinary, unnormalized trace of matrices whose indices $a,b$ label the retained observables $T_a$: it sums their diagonal entries. The last term accounts for a quadratic phase and is nonnegative. We compute $C_0$, $\Omega$, and $R$ once for each planar state, then minimize \eqref{eqn:auxiliary-energy} over $\boldsymbol\eta$. The covariances follow from \eqref{eqn:auxiliary-covariance}; they are not fitted. Varying every real entry of $B_{\boldsymbol\eta}$ independently gives $C_{\boldsymbol\eta}RC_{\boldsymbol\eta}=\Omega/4$, a benchmark for the short auxiliary families.

For the short two-matrix auxiliary families below, we expand the logarithm of the determinant and keep quadratic terms in the selected trace fluctuations to obtain $B_{\boldsymbol\eta}$ as a function of the couplings. For each trial choice, we evaluate the covariance using \eqref{eqn:auxiliary-covariance} and the energy using \eqref{eqn:auxiliary-energy}, rejecting choices for which the covariance is not positive definite. We minimize the energy by comparing nearby coupling values, repeating the search from several starting points and retaining the lowest value found. With $C_0$, $\Omega$ and $R$ fixed, these evaluations require no further SD solutions or matrix integrals. The construction of these matrices and the minimization procedure are detailed in \S\ref{app:auxiliary-optimization}.

This calculation depends on the accuracy of the planar state and the retained fluctuation basis, whose convergence must be checked alongside $L$ and $K$. We require $C_0$ and $C_{\boldsymbol\eta}$ to be positive definite. This ensures local Gaussian convergence; a finite-$N$ wavefunction additionally requires convergence of the original auxiliary integral and its norm.

\section{Numerical Results}

Our variational ans\"atze converge remarkably quickly, and precisely match existing published data to many digits. For covariances we could not find results to match to, so we perform our own independent checks through resummed quantum perturbation theory (Appendix \ref{app:resummation}) and Monte Carlo (Appendix \ref{app:monte-carlo}). We report the results below for several different systems (single matrix with strong quartic coupling, two-matrix at moderate and strong commutator-squared coupling, and nine-matrix bosonic BFSS). All numerics (besides the Monte Carlo, which needed more time) were done in at most a few hours via Python scripts running on a 16-core 2024 Dell Inspiron 16 -- no additional computational resources were needed for the precision achieved.

\subsection{Energies and Single-Trace Expectation Values}

\subsubsection{One Matrix}\label{sec:one-matrix}

We first consider a single Hermitian matrix with a quartic interaction,
\begin{equation}\label{eqn:one-matrix-model}
H=\frac{1}{2N}\Tr\Pi^2+N\Tr V(X),
\qquad V(x)=\frac12x^2+\lambda x^4.
\end{equation}
We take real coefficients and retain the even words appropriate to the positive, parity-even ground state:
\begin{equation}\label{eqn:one-matrix-ansatz}
f_{\boldsymbol g}(X)=N\sum_{k=1}^{L/2}g_{X^{2k}}X^{2k}.
\end{equation}
Thus the cutoff $L$ gives $L/2$ variational parameters, all of which are reoptimized at each cutoff.

The exact planar reference follows from the equivalence of the singlet sector to $N$ noninteracting fermions: diagonalizing $X$ and absorbing the Vandermonde determinant into the wavefunction makes the eigenvalue wavefunction antisymmetric \cite{Brezin:1978Planar}. Filling the lowest single-particle levels gives the large-$N$ eigenvalue density
\begin{equation}\label{eqn:one-matrix-reference-density}
\rho_*(x)=\frac1\pi\sqrt{A^2-x^2}\sqrt{1+2\lambda(A^2+x^2)},
\qquad \int_{-A}^{A}\rho_*(x)\,dx=1,
\end{equation}
on the interval $[-A,A]$, and zero outside. We determine $A$ from the normalization and compute the exact planar ground-state energy per $N^2$ and normalized single-trace moments by numerical integration:
\begin{equation}\label{eqn:one-matrix-reference-observables}
e_*=\int_{-A}^{A}\left[\frac{\pi^2}{6}\rho_*^3(x)+V(x)\rho_*(x)\right]dx,
\qquad
m_{2k,*}=\int_{-A}^{A}x^{2k}\rho_*(x)\,dx.
\end{equation}
These references are independent of the variational ansatz and refer to the planar limit, rather than finite $N$.

For this model we can also evaluate the trial matrix integral exactly at large $N$. Its single-interval density is
\begin{equation}\label{eqn:one-matrix-trial-density}
\rho_L(x)=\frac1\pi Q_{L-2}(x)\sqrt{a^2-x^2},
\qquad \int_{-a}^{a}\rho_L(x)\,dx=1,
\end{equation}
where $Q_{L-2}$ is an even polynomial of degree at most $L-2$. The planar variational energy and moments are given by \eqref{eqn:one-matrix-reference-observables} with $\rho_*$ and $A$ replaced by $\rho_L$ and $a$. This parametrizes the same trial states as varying the exponent coefficients. The figures therefore isolate the error due to $L$; no cumulant truncation $K$ is used here. We display only normalizable states.\footnote{The displayed cutoffs are $2,4,8,12,16,20$. The unconstrained solutions at $L=6,10,14,18$ have negative leading exponent coefficients and are excluded.}

\begin{figure}[tbp]
\centering
\includegraphics[width=0.82\textwidth]{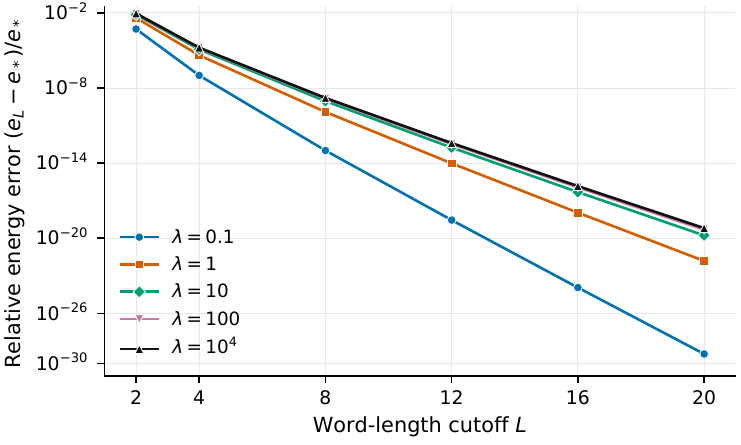}
\caption{Relative error of the optimized planar energy versus word cutoff $L$, for five quartic couplings. The vertical axis is the excess energy $e_L-e_*$ divided by the exact planar energy $e_*$ from \eqref{eqn:one-matrix-reference-observables}, on a logarithmic scale. Every coefficient is reoptimized at each plotted cutoff. Only normalizable states are shown.}
\label{fig:one-matrix-energy}
\end{figure}

\begin{figure}[tbp]
\centering
\includegraphics[width=\textwidth]{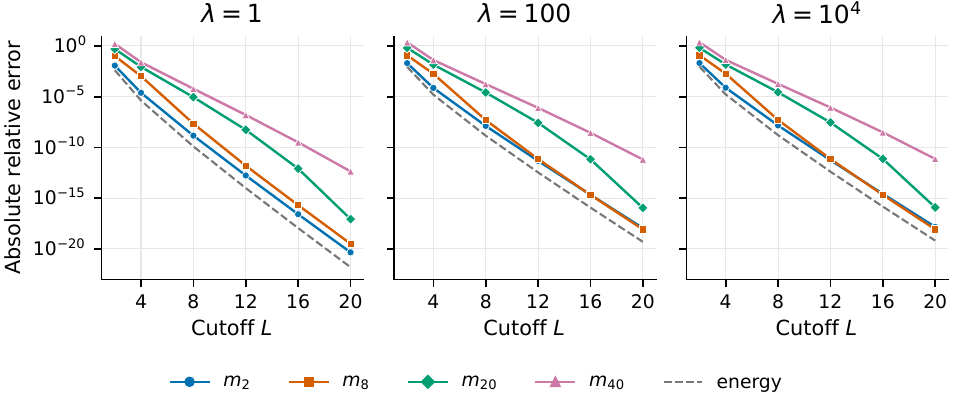}
\caption{Absolute relative errors of the normalized moments of $X^2$, $X^8$, $X^{20}$ and $X^{40}$ in the energy-minimized states, versus word cutoff $L$. The three panels show couplings $1$, $100$ and $10^4$. Each error is divided by the corresponding exact Fermi-sea moment in \eqref{eqn:one-matrix-reference-observables}; the vertical scale is logarithmic. Dashed curves reproduce the relative energy errors for comparison. All moments are predictions of the optimized states, not additional fitting targets.}
\label{fig:one-matrix-moments}
\end{figure}

Figure~\ref{fig:one-matrix-energy} shows rapid convergence of the optimized energy $e_L$ throughout the coupling range. Figure~\ref{fig:one-matrix-moments} tests the same states on moments through $X^{40}$, none of which are fitted. These longer words are a more demanding test: at coupling $10^4$, two parameters give relative errors of $1.7\times10^{-5}$ in the energy and $4.2\times10^{-2}$ in the $X^{40}$ moment, while four parameters reduce these to $1.6\times10^{-9}$ and $1.8\times10^{-4}$. With ten parameters the $X^{40}$ error is $7.3\times10^{-12}$, even though this word is twice as long as the longest word in the exponent.

For a concrete example, the ten-coefficient optimum at $\lambda=100$ ($L=20$) is
\begin{equation}\label{eqn:one-matrix-explicit-state}
\Psi_{20}(X)\propto\exp\!\left[-N\Tr s_{20}(Z)\right],
\qquad Z=100^{1/6}X,
\end{equation}
where, rounding the coefficients to six significant figures,
\begin{equation}\label{eqn:one-matrix-explicit-polynomial}
\begin{aligned}
s_{20}(Z)\simeq{}&0.591822 Z^2+0.179995 Z^4-2.76033\times10^{-2}Z^6\\
&+9.04847\times10^{-3}Z^8-3.84912\times10^{-3}Z^{10}
+1.81232\times10^{-3}Z^{12}\\
&-8.23860\times10^{-4}Z^{14}+3.08393\times10^{-4}Z^{16}
-7.82303\times10^{-5}Z^{18}\\
&+9.59730\times10^{-6}Z^{20}.
\end{aligned}
\end{equation}
The rescaling removes the overall quartic-coupling scale. The coefficient magnitudes decrease rapidly in these units; they need not decrease in the original matrix units. The convergence tests above use the unrounded coefficients.

\subsubsection{Two Matrices}\label{sec:two-matrices}

We next consider two traceless Hermitian matrices with a commutator interaction~\cite{Rodrigues:2010StrongCoupling,Mathaba:2024MasterField},
\begin{equation}\label{eqn:two-matrix-model}
H=\frac{1}{2N}\Tr\!\left(\Pi_X^2+\Pi_Y^2\right)
+\frac N2\Tr\!\left[\mu(X^2+Y^2)-[X,Y]^2\right].
\end{equation}
The quartic coupling has been set to one by rescaling, and $\mu$ is the mass squared in these units. We study the massive case $\mu=1$ and the massless case $\mu=0$, and write $e$ for the planar energy per $N^2$. We take real exponent coefficients and impose invariance under reversal of words, interchange of $X$ and $Y$, and independent sign flips of either matrix. After identifying cyclic duplicates, the complete ansatz through lengths $2$, $4$, $6$ and $8$ has $1$, $4$, $8$ and $20$ free coefficients, respectively. We optimize these coefficients using the truncated SD equations of \S2.1.\footnote{Some optimized polynomial exponents are not globally normalizable as written: for example, the massive $L=8$ solution has a negative coefficient of $X^8$ along $Y=0$. The numbers reported here are estimates from the truncated planar SD equations, not certified energy upper bounds for normalizable finite-$N$ wavefunctions.}

At $\mu=1$, the twenty-coefficient optimum of the truncated SD calculation, with $L=8$ and $K=12$, takes the explicit form
\begin{equation}\label{eqn:two-matrix-explicit-state}
\Psi_8(X,Y)\propto\exp\!\left[-N\Tr\mathcal S p_8(X,Y)\right].
\end{equation}
Under the trace, $\mathcal S$ sums the distinct cyclic words obtained by reversing a word and interchanging $X$ and $Y$, counting each once. For example,
\begin{equation}\label{eqn:two-matrix-word-sum}
\mathcal S(X^2)=X^2+Y^2,
\qquad \mathcal S(X^2Y^2)=X^2Y^2.
\end{equation}
The polynomial, again rounded to six significant figures, is
\begin{equation}\label{eqn:two-matrix-explicit-polynomial}
\begin{aligned}
p_8(X,Y)\simeq{}&0.583894 X^2-7.44907\times10^{-4}X^4\\
&+0.197115 X^2Y^2-0.198605 XYXY\\
&+1.45663\times10^{-5}X^6-3.88667\times10^{-3}X^4Y^2\\
&+1.77199\times10^{-2}X^3YXY-1.37893\times10^{-2}X^2YX^2Y\\
&-4.17073\times10^{-8}X^8+2.97867\times10^{-5}X^6Y^2\\
&-4.27939\times10^{-4}X^5YXY+1.39845\times10^{-3}X^4YX^2Y\\
&+8.60012\times10^{-5}X^4Y^4-1.00039\times10^{-3}X^3YX^3Y\\
&-8.18777\times10^{-4}X^3YXY^3+1.09692\times10^{-3}X^3Y^2XY^2\\
&-2.05318\times10^{-3}X^2YXYXY^2+2.68891\times10^{-3}X^2YXY^2XY\\
&+5.26465\times10^{-5}X^2Y^2X^2Y^2+7.21613\times10^{-4}(XY)^4.
\end{aligned}
\end{equation}
The largest coefficient magnitude at each successive length is $0.584$, $0.199$, $0.0177$ and $0.00269$. This is the same planar SD solution used in table~\ref{tab:two-matrix-bootstrap}, with the normalizability qualification stated above.

\begin{figure}[tbp]
\centering
\begin{subfigure}[t]{0.485\textwidth}
\centering
\includegraphics[width=\linewidth]{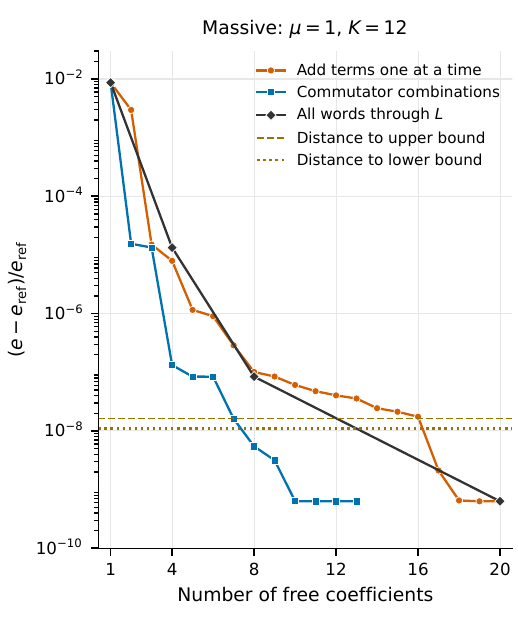}
\caption{Comparison with our best computed state.}
\label{fig:two-matrix-energy}
\end{subfigure}\hfill
\begin{subfigure}[t]{0.485\textwidth}
\centering
\includegraphics[width=\linewidth]{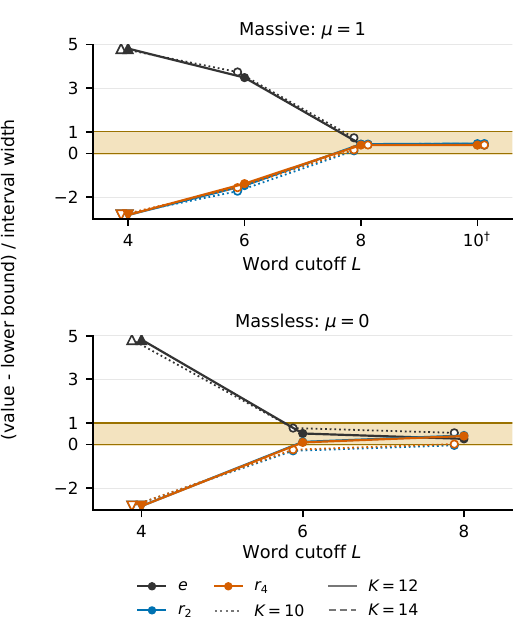}
\caption{Comparison with bootstrap intervals.}
\label{fig:two-matrix-bootstrap}
\end{subfigure}
\caption{Two-matrix results. \textbf{(a)} Relative excess SD energy above $e_{\mathrm{ref}}$ in \eqref{eqn:two-matrix-internal-reference}, versus the number of fitted coefficients, at $\mu=1$ and $K=12$. The reference is our best computed estimate, not an exact ground-state energy. The dashed and dotted horizontal lines show its relative distances to the upper and lower bootstrap bounds; they do not enclose an allowed band on this axis. \textbf{(b)} Predictions for $e$, $r_2$ and $r_4$ versus word cutoff $L$. Each vertical coordinate is the prediction minus its lower bootstrap bound, divided by the interval width, so the shaded band from zero to one denotes agreement with that bound. Colour specifies the observable and line style the cumulant cutoff $K$; filled markers indicate $K=12$. Triangles denote values outside the plotted range, listed in table~\ref{tab:two-matrix-bootstrap}. The $L=10^\dagger$ points use the restricted seven-coefficient exponent.}
\label{fig:two-matrix-comparison}
\end{figure}

Figure~\ref{fig:two-matrix-energy} compares three ways of enlarging the exponent: adding symmetry-related sums of words one at a time, adding fixed combinations built from commutators, and including all allowed words through $L$. Every retained coefficient is refitted at each step. The commutator combinations respect rotations and reflections of $(X,Y)$ and begin, in order, with
\begin{equation}\label{eqn:two-matrix-short-terms}
\begin{gathered}
X^2+Y^2,\qquad [X,Y]^2,\\
X^4+Y^4+X^2Y^2+XYXY,\qquad
[X,[X,Y]]^2+[Y,[X,Y]]^2.
\end{gathered}
\end{equation}
Each displayed polynomial carries one independent coefficient under the trace. Further terms complete the thirteen independent rotation-invariant combinations through length eight. At $K=12$, four coefficients give an energy within $1.3\times10^{-7}$ of our best computed estimate, in relative units. All thirteen reproduce the twenty-coefficient energy to numerical precision.

The reference for this comparison is our best available approximation to the optimal state, obtained with seven coefficients multiplying fixed combinations of words through length ten, specified in \S\ref{app:single-trace-optimization}. Its SD energy at $K=12$ is
\begin{equation}\label{eqn:two-matrix-internal-reference}
e_{\mathrm{ref}}=1.1720983888828.
\end{equation}
We have no independently known exact wavefunction or energy to use in this panel. Its vertical axis therefore measures convergence towards our best computed state, not error relative to the exact ground state; neither global optimality nor the displayed digits of $e_{\mathrm{ref}}$ are established.

The crucial independent test comes from the precision bootstrap of Lin and Zheng~\cite{Lin:2026Precision}. They combine canonical commutation relations, stationarity, large-$N$ factorization and positivity to bound the planar ground-state energy and two moments. To compare with their normalization, we define
\begin{equation}\label{eqn:two-matrix-bootstrap-moments}
r_2=m_{X^2}+m_{Y^2},\qquad
r_4=\frac14\left(m_{X^4}+m_{Y^4}+2m_{XYXY}\right).
\end{equation}
Table~\ref{tab:two-matrix-bootstrap} lists our predictions and their level-14 bounds; their bootstrap level is a separate cutoff from our $L$ and $K$. Figure~\ref{fig:two-matrix-bootstrap} places each prediction within its corresponding interval. The moments are computed from the energy-optimized coefficients and are not fitted to the bootstrap results.

For the massive model, all three predictions at $L=8$ lie inside the bootstrap intervals for $K=10$, $12$ and $14$, whereas $L=6$ lies outside. For the massless model, both $L=6$ and $L=8$ lie inside all three intervals at $K=12$. Cumulant convergence matters here: at $L=8$, increasing $K$ from $10$ to $12$ shifts $r_2$ by about half the bootstrap interval's width, moving it inside the interval. These comparisons support the accuracy of a short exponent, while the finite interval widths limit the precision that can be independently verified.

\begin{table}[tbp]
\centering\small
\setlength{\tabcolsep}{3pt}
\begin{tabular}{ccccrrr}
\toprule
$\mu$ & $L$ & coefficients & $K$ & $e$ & $r_2$ & $r_4$ \\
\midrule
1 & 4 & 4 & 8 & 1.1721141609 & 0.7779346783 & 0.1584641553 \\
 & 4 & 4 & 10 & 1.1721140575 & 0.7779356471 & 0.1584645242 \\
 & 4 & 4 & 12 & 1.1721140551 & 0.7779356747 & 0.1584645336 \\
 & 6 & 8 & 10 & 1.1720984958 & 0.7780083579 & 0.1585055827 \\
 & 6 & 8 & 12 & 1.1720984876 & 0.7780084513 & 0.1585056177 \\
 & 8 & 20 & 10 & \textbf{1.1720983992} & \textbf{0.7780090265} & \textbf{0.1585059126} \\
 & 8 & 20 & 12 & \textbf{1.1720983896} & \textbf{0.7780091349} & \textbf{0.1585059543} \\
 & 8 & 20 & $14^{\ddagger}$ & \textbf{1.1720983892} & \textbf{0.7780091360} & \textbf{0.1585059549} \\
 & $10^{\dagger}$ & 7 & 12 & \textbf{1.1720983889} & \textbf{0.7780091431} & \textbf{0.1585059543} \\
 & $10^{\dagger}$ & 7 & $14^{\ddagger}$ & \textbf{1.1720983884} & \textbf{0.7780091448} & \textbf{0.1585059553} \\
\cmidrule(l){2-7}
 & \multicolumn{3}{l}{Bootstrap lower bound} & 1.172098376 & 0.77800898 & 0.15850588 \\
 & \multicolumn{3}{l}{Bootstrap upper bound} & 1.172098408 & 0.77800934 & 0.15850607 \\
\midrule
0 & 4 & 4 & 10 & 0.7081725 & 1.1515363 & 0.3684282 \\
 & 4 & 4 & 12 & 0.7081692 & 1.1515992 & 0.3684660 \\
 & 6 & 8 & 10 & \textbf{0.7078594} & 1.1540868 & 0.3704802 \\
 & 6 & 8 & 12 & \textbf{0.7078504} & \textbf{1.1542502} & \textbf{0.3705809} \\
 & 8 & 20 & 10 & \textbf{0.7078515} & 1.1541869 & \textbf{0.3705572} \\
 & 8 & 20 & 12 & \textbf{0.7078415} & \textbf{1.1543676} & \textbf{0.3706686} \\
\cmidrule(l){2-7}
 & \multicolumn{3}{l}{Bootstrap lower bound} & 0.707832 & 1.15420 & 0.37055 \\
 & \multicolumn{3}{l}{Bootstrap upper bound} & 0.707868 & 1.15460 & 0.37085 \\
\bottomrule
\end{tabular}

\caption{Planar SD estimates compared with the precision bootstrap bounds of Lin and Zheng~\cite{Lin:2026Precision}. The mass squared is $\mu$; $L$ is the longest word retained in the exponent, and $K$ is the cumulant cutoff. The coefficient column counts independently fitted parameters. Bold entries lie within the corresponding bootstrap interval. $\dagger$: seven coefficients multiplying fixed combinations of words through length ten, rather than the complete length-ten ansatz. $\ddagger$: the $K=12$ coefficients reevaluated at $K=14$, without further optimization. Other rows are optimized at the stated $K$. Extra digits are shown to resolve changes with the cutoffs, not as a claim of independently verified accuracy.}
\label{tab:two-matrix-bootstrap}
\end{table}
\FloatBarrier

\subsubsection{$D=9$ Bosonic BFSS}\label{sec:bosonic-bfss}

Finally, we consider the bosonic part of the Banks--Fischler--Shenker--Susskind (BFSS) model~\cite{Banks:1997BFSS}, with nine traceless Hermitian matrices and unit 't Hooft coupling,
\begin{equation}\label{eqn:bosonic-bfss-model}
H=\frac{1}{2N}\sum_{i=1}^{9}\Tr\Pi_i^2
-\frac N4\sum_{i,j=1}^{9}\Tr[X^i,X^j]^2.
\end{equation}
We take a real exponent invariant under orthogonal rotations and reflections of the nine matrices, and under reversal of words. This symmetry allows us to work with contractions of matrix labels instead of enumerating words in nine distinct letters.

Orthogonal symmetry allows each free cumulant of even word length $n$ to be written as a sum of pairwise index contractions~\cite{Brauer:1937Invariants,Lehrer:2012Orthogonal}:
\begin{equation}\label{eqn:chord-cumulants}
\kappa_{i_1\cdots i_n}
=\sum_{\pi\in\mathcal P_2(n)}k_\pi
\prod_{\{a,b\}\in\pi}\delta_{i_a i_b}.
\end{equation}
Here $\mathcal P_2(n)$ is the set of pairings of the $n$ positions, $\delta$ is the Kronecker delta, and $k_\pi$ are scalar coefficients determined by the SD equations. Placing the positions around a circle and joining each pair gives a \emph{chord diagram}. Cyclicity and reversal identify diagrams related by rotation or reflection. Crossing chords are retained: they contract matrix labels, whereas the noncrossing partitions in \eqref{eqn:free-moment-cumulant} organize the reconstruction of moments from cumulants. Summing contracted indices gives a factor of nine for each closed index loop, reducing the SD calculation to operations on these diagrams. There are $1$, $2$, $5$, $17$ and $79$ independent coefficients at lengths $2$, $4$, $6$, $8$ and $10$, respectively, so $K=8$ and $K=10$ require only $25$ and $104$ cumulant unknowns. The pairing representation itself introduces no approximation at these lengths.

The same contractions organize the exponent. Through length four, the three independent polynomials are
\begin{equation}\label{eqn:bfss-quartic-ansatz}
\sum_i X^iX^i,\qquad
\sum_{i,j}X^iX^iX^jX^j,\qquad
\sum_{i,j}X^iX^jX^iX^j,
\end{equation}
with one coefficient for each polynomial under the trace; every index is summed from $1$ to $9$. The five additional contractions at length six give eight coefficients in total. We reoptimize all retained exponent coefficients at each pair of cutoffs $L$ and $K$, using the SD solution and energy minimization of \S2.1.\footnote{The optimized polynomial exponents have unstable directions at both $L=4$ and $L=6$: one nonzero matrix suffices at $L=4$, and two noncommuting matrices suffice at $L=6$. As in \S\ref{sec:two-matrices}, we report truncated planar SD estimates, not certified variational upper bounds for normalizable finite-$N$ wavefunctions.}

\begin{figure}[tbp]
\centering
\includegraphics[width=\textwidth]{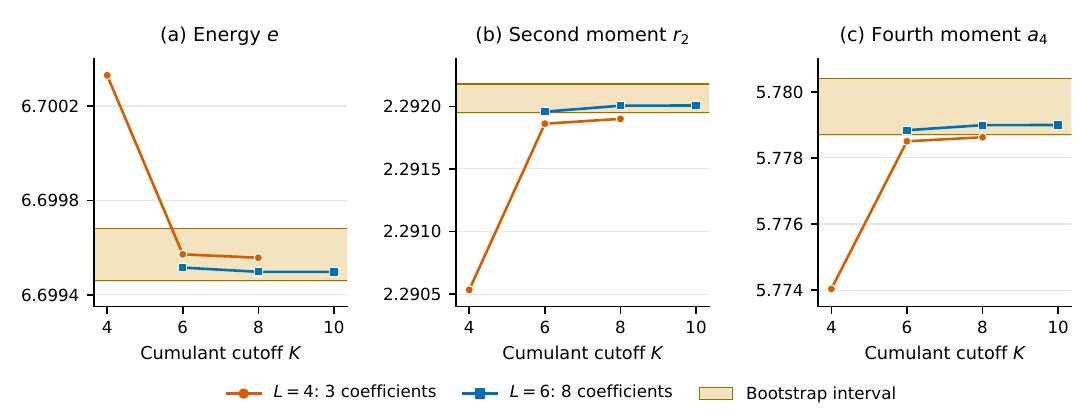}
\caption{Bosonic BFSS with nine matrices: the planar energy $e$ and moments $r_2$ and $a_4$, defined in \eqref{eqn:bfss-observables}, versus cumulant cutoff $K$. Orange circles use the complete invariant exponent through length four (three coefficients); blue squares include length six (eight coefficients). All exponent coefficients are reoptimized at each point. Shading shows the independent bootstrap intervals~\cite{Lin:2026Precision}. Table~\ref{tab:bfss-bootstrap} resolves the small changes between the last points.}
\label{fig:bfss-bootstrap}
\end{figure}

We compare the energy per $N^2$ and the following moments:
\begin{equation}\label{eqn:bfss-observables}
\begin{aligned}
r_2&=\sum_{i=1}^{9}m_{X^iX^i},\\
a_4&=\sum_{i,j=1}^{9}m_{X^iX^iX^jX^j},\qquad
c_4=\sum_{i,j=1}^{9}m_{X^iX^jX^iX^j}.
\end{aligned}
\end{equation}
The two fourth-order moments differ in the ordering of the matrices. Figure~\ref{fig:bfss-bootstrap} and table~\ref{tab:bfss-bootstrap} compare $e$, $r_2$ and $a_4$ with the independent level-11 bootstrap intervals of Lin and Zheng~\cite{Lin:2026Precision}; this bootstrap level is distinct from our cutoffs $L$ and $K$. At $L=4$ and $K=8$, the energy lies inside its interval, but both moments lie below theirs. With eight coefficients ($L=6$), all three predictions lie inside their intervals at $K=6$, $8$ and $10$. The moments are predictions of the energy-minimized solution and are not fitted to the bounds. We also report $c_4$, for which their table gives no separate interval.

Increasing $K$ from $6$ to $8$, and then to $10$, changes the length-six energy by $1.8\times10^{-5}$ and $5.1\times10^{-7}$, respectively. These shifts test the cumulant approximation. Increasing $L$ from $4$ to $6$ at fixed $K=8$ instead lowers the energy by $6.0\times10^{-5}$. A controlled length-eight calculation is still needed to assess the remaining error from the exponent, so the observed convergence with $K$ does not justify error bars significantly narrower than the bootstrap intervals.

\begin{table}[tbp]
\centering\small
\setlength{\tabcolsep}{4pt}
\begin{tabular}{cccrrrr}
\toprule
$L$ & coefficients & $K$ & $e$ & $r_2$ & $a_4$ & $c_4$ \\
\midrule
4 & 3 & 4 & 6.7003299 & 2.2905340 & 5.7740342 & 1.3071465 \\
4 & 3 & 6 & \textbf{6.6995715} & 2.2918614 & 5.7785043 & 1.3121231 \\
4 & 3 & 8 & \textbf{6.6995570} & 2.2919001 & 5.7786250 & 1.3122536 \\
6 & 8 & 6 & \textbf{6.6995153} & \textbf{2.2919575} & \textbf{5.7788385} & 1.3124950 \\
6 & 8 & 8 & \textbf{6.6994974} & \textbf{2.2920052} & \textbf{5.7789913} & 1.3126597 \\
6 & 8 & 10 & \textbf{6.6994969} & \textbf{2.2920071} & \textbf{5.7789968} & 1.3126656 \\
\midrule
\multicolumn{3}{l}{Bootstrap lower bound} & 6.69946 & 2.29195 & 5.7787 & --- \\
\multicolumn{3}{l}{Bootstrap upper bound} & 6.69968 & 2.29218 & 5.7804 & --- \\
\bottomrule
\end{tabular}

\caption{Nine-matrix planar SD estimates and the level-11 bootstrap bounds~\cite{Lin:2026Precision}. $L$ specifies the exponent, $K$ specifies the cumulants retained in its evaluation, and the coefficient column counts fitted exponent parameters. Bold values lie inside their corresponding bootstrap intervals. The moment $c_4$ is defined in \eqref{eqn:bfss-observables}; a dash indicates that no separate bound is quoted in the reference. Digits beyond the interval widths resolve numerical changes with the cutoffs and are not independently verified.}
\label{tab:bfss-bootstrap}
\end{table}
\FloatBarrier

\subsection{Connected Two-Point Functions}\label{sec:connected-covariances}

We now test the leading covariances of single-trace operators. For one matrix, we minimize the full finite-$N$ energy and compare the predicted covariances with exact free-fermion results. For two matrices, we use the procedure of \S2.2: holding the optimized planar moments fixed, we minimize the $O(N^0)$ energy over the auxiliary couplings. For the two-matrix models that can't be solved exactly we compare these predictions with quantum perturbation theory resummed using Pad\'e approximants and ground-state projection Monte Carlo estimates extrapolated to large $N$, subject to the normalizability caveat explained in \S\ref{sec:two-matrix-covariances}. In both examples the covariances are predictions, rather than fitted data.

\subsubsection{One Matrix}\label{sec:one-matrix-covariances}

\begin{figure}[tbp]
\centering
\includegraphics[width=\textwidth]{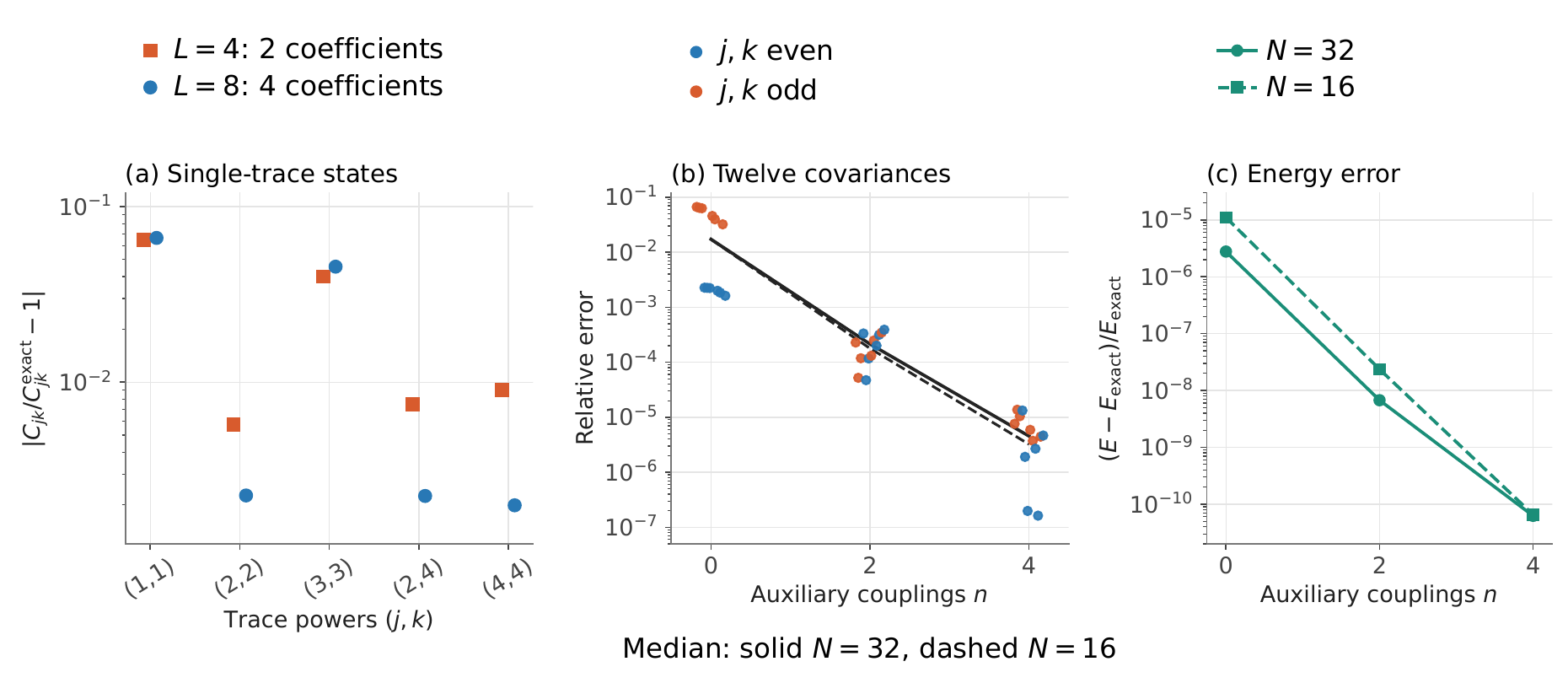}
\caption{Covariances in the quartic one-matrix model at $\lambda=1$, compared with numerically converged exact free-fermion results at the same $N$. \textbf{(a)} Absolute relative errors of five covariances at $N=32$ for the two- and four-coefficient single-trace states; the horizontal labels specify the powers in \eqref{eqn:one-matrix-connected-covariance}. \textbf{(b)} Absolute relative errors of all twelve covariances for \eqref{eqn:one-matrix-covariance-auxiliary}, versus the number $n$ of fitted coefficients in $u$. The same $n$ parameters specify the polynomial auxiliary action, with words through length $2n+2$. Points show $N=32$, with small horizontal offsets for visibility; blue and orange distinguish even and odd powers. Black lines show the median at $N=32$ (solid) and $N=16$ (dashed). \textbf{(c)} Energy excess divided by the exact finite-$N$ ground-state energy at the same $N$. There are $4+n$ fitted parameters in (b,c), and $n=0$ is the $L=8$ single-trace state. The saved optimized coefficients are reevaluated with a refined reference and direct integration.}
\label{fig:one-matrix-covariances}
\end{figure}

For the quartic model \eqref{eqn:one-matrix-model} at $\lambda=1$, we test the twelve nonzero covariances
\begin{equation}\label{eqn:one-matrix-connected-covariance}
C_{jk}=\langle\Tr X^j\Tr X^k\rangle-\langle\Tr X^j\rangle\langle\Tr X^k\rangle,
\qquad 1\leq j\leq k\leq6,\quad j+k\ \text{even}.
\end{equation}
Mixed-parity covariances vanish. We retain the trace mode and work in the $U(N)$ singlet sector, whose exact ground state is obtained by filling the lowest $N$ levels of the one-particle quartic oscillator~\cite{Brezin:1978Planar}. Numerical diagonalization gives independent reference covariances at the same finite $N$ as each trial state. Figure~\ref{fig:one-matrix-covariances}(a) shows that increasing the single-trace cutoff from $L=4$ to $L=8$ improves the displayed even-power covariances, but leaves a relative error of $6.6\times10^{-2}$ in $\mathrm{Var}(\Tr X)$ at $N=32$. Accurate planar moments therefore do not by themselves guarantee accurate covariances.

Following our hypothesis, we test how auxiliary variables as in \eqref{eqn:basic-auxiliary-ansatz} can improve the behavior of the covariances. For the one matrix case, the following construction works quite well. First, we choose
\begin{equation}\label{eqn:one-matrix-auxiliary-polynomials}
s(x)=\sum_{r=1}^{4}g_{X^{2r}}x^{2r},
\qquad u(x)=x+\sum_{r=1}^{n}c_r x^{2r+1}.
\end{equation}
Here $n$ counts the fitted coefficients $c_r$. The linear map $\mathcal J_u$ is the matrix derivative of $u$, acting on any matrix $Z$ as
\begin{equation}\label{eqn:one-matrix-auxiliary-kernel}
\begin{aligned}
\mathcal J_u(X)[Z]
&=\left.\frac{d}{d\epsilon}u(X+\epsilon Z)\right|_{\epsilon=0}\\
&=Z+\sum_{r=1}^{n}c_r\sum_{k=0}^{2r}X^kZX^{2r-k}.
\end{aligned}
\end{equation}

\begin{samepage}
Introduce one Hermitian auxiliary matrix $A$ and two independent Grassmann matrices $\chi,\bar\chi$, each with $N^2$ anticommuting components. We take the restricted auxiliary ansatz
\begin{equation}\label{eqn:one-matrix-auxiliary-integral}
\begin{aligned}
\Psi_n(X)&\propto e^{-N\Tr s(X)}\int dA\,d\bar\chi\,d\chi\,e^{-\Tr h_u(X,A,\bar\chi,\chi)},\\
h_u(X,A,\bar\chi,\chi)&=N\left[\frac12 A\,\mathcal J_u(X)[A]+\bar\chi\,\mathcal J_u(X)[\chi]\right].
\end{aligned}
\end{equation}
\end{samepage}
The measures integrate over the $N^2$ real components of $A$ and the $2N^2$ independent Grassmann components. Each $c_r$ multiplies a fixed sum of words of length $2r+2$ in $h_u$. We require $u'>0$, which makes $\mathcal J_u$ positive and ensures convergence of the ordinary auxiliary integral.

Ordinary Gaussian integration gives an inverse square root of the determinant of $\mathcal J_u$ on the $N^2$-dimensional matrix space, while the Grassmann pair gives the determinant itself~\cite{Berezin:1966SecondQuantization}. Writing $x_i$ for the eigenvalues of $X$, integrating out the auxiliaries therefore gives
\begin{equation}\label{eqn:one-matrix-covariance-auxiliary}
\begin{aligned}
\Psi_n(X)&=\mathcal N e^{-N\Tr s(X)}[\det\mathcal J_u(X)]^{1/2}\\
&=\mathcal N e^{-N\Tr s(X)}
\prod_{i=1}^{N}\sqrt{u'(x_i)}
\prod_{i<j}\frac{u(x_i)-u(x_j)}{x_i-x_j}.
\end{aligned}
\end{equation}
Here $\mathcal N$ normalizes the state, and coincident eigenvalues are understood by continuity.

Multiplying the wavefunction by the Vandermonde determinant gives a Slater determinant, allowing direct finite-$N$ evaluation of the energy and covariances. We jointly minimize the full finite-$N$ energy over the four coefficients in $s$ and the $n$ coefficients in $u$, separately at each $N$; no cumulant truncation is used. The displayed optimized maps have $n=0,2,4$ and are increasing on the entire real axis.\footnote{The unconstrained fits at $n=1,3,5$ have zeros of $u'$ and are excluded. The positive leading coefficient of $s$ ensures normalizability of the displayed wavefunctions.}

Figure~\ref{fig:one-matrix-covariances}(b,c) shows the improvement with auxiliaries, which we generate by Taylor expanding $u(x)$. At $N=32$, the median relative covariance error falls from $1.7\times10^{-2}$ at $n=0$ to $2.1\times10^{-4}$ at $n=2$ and $4.6\times10^{-6}$ at $n=4$; the corresponding worst errors are $6.6\times10^{-2}$, $3.9\times10^{-4}$ and $1.4\times10^{-5}$. The relative energy error falls from $2.8\times10^{-6}$ to $6.7\times10^{-9}$ and $6.2\times10^{-11}$. The same improvement occurs at $N=16$. These are finite-$N$ comparisons of energy-minimized states, with every covariance predicted rather than fitted.

\FloatBarrier

\subsubsection{Two Matrices}\label{sec:two-matrix-covariances}

For the two-matrix model we fix the oscillator mass to one and vary the commutator coupling:
\begin{equation}\label{eqn:two-matrix-covariance-model}
H_\lambda=\frac{1}{2N}\Tr(\Pi_X^2+\Pi_Y^2)
+\frac N2\Tr(X^2+Y^2)-\frac{N\lambda}{2}\Tr[X,Y]^2.
\end{equation}
This is the model of \S\ref{sec:two-matrices} in different units. For positive coupling, the conversion to unit quartic coupling is
\begin{equation}\label{eqn:two-matrix-covariance-scaling}
X=\lambda^{-1/6}\widetilde X,\qquad
Y=\lambda^{-1/6}\widetilde Y,\qquad
\mu=\lambda^{-2/3}.
\end{equation}
The massless model is therefore reached at infinite coupling in the rescaled variables.

We consider all ten leading covariances of four real trace observables:
\begin{equation}\label{eqn:two-matrix-connected-covariances}
\begin{gathered}
\begin{aligned}
T_1&=\Tr(X^2+Y^2),& T_2&=\Tr(X^4+Y^4),\\
T_3&=\Tr(X^2Y^2),& T_4&=\Tr(XYXY),
\end{aligned}
\\
C_{ab}=\lim_{N\to\infty}\bigl(\langle T_aT_b\rangle-\langle T_a\rangle\langle T_b\rangle\bigr),
\qquad 1\leq a\leq b\leq4.
\end{gathered}
\end{equation}
The traces are unnormalized. We use the twenty-coefficient planar exponent at $L=8$, $K=12$. Keeping its planar moments fixed, we minimize the $O(N^0)$ energy over one or three auxiliary couplings, or over all ten real quadratic coefficients on these four observables.

The short families use one Gaussian auxiliary matrix $A$, with quadratic operators
\begin{equation}\label{eqn:two-matrix-short-auxiliaries}
\begin{aligned}
\mathcal M_1[A]&=A+\ell\bigl([X,[X,A]]+[Y,[Y,A]]\bigr),\\
\mathcal M_3[A]&=A+a\{X^2+Y^2,A\}+b(XAX+YAY)\\
&\quad+c[[X,Y],[[X,Y],A]].
\end{aligned}
\end{equation}
Here braces denote the anticommutator; $\ell$ is one fitted coupling, and $a,b,c$ are three. We expand the auxiliary determinant and retain only quadratic products of the four trace fluctuations in \eqref{eqn:two-matrix-connected-covariances}. The comparison therefore tests these projected Gaussian corrections, rather than the complete determinants. As in \S2.2, positive fluctuation covariances ensure only local convergence; global normalizability of the underlying states remains to be established.

Our independent references come from quantum perturbation theory (QPT) through order $\lambda^{10}$, resummed using Pad\'e approximants. At $\lambda=1$, table~\ref{tab:two-matrix-covariances} uses the median of three $[5/5]$ approximants after the conformal change of variable
\begin{equation}\label{eqn:covariance-pade-map}
w=\frac{\sqrt{1+\lambda/\rho}-1}{\sqrt{1+\lambda/\rho}+1},
\qquad \rho\in\{0.19,0.21,0.23\}.
\end{equation}
The map sends the complex coupling plane cut along $(-\infty,-\rho]$ to the unit disk, placing all positive couplings inside. Reexpanding in $w$ before constructing Pad\'e approximants incorporates this assumed singularity structure and can improve continuation beyond the original convergence radius~\cite{Costin:2020PhysicalResurgentExtrapolation}. The chosen $\rho$ values bracket an empirical estimate from negative-axis poles of the energy Pad\'e approximants; their variation tests sensitivity to the assumed cut.

The notation $[p/q]$ specifies numerator and denominator degrees. Even without auxiliary corrections, the single-trace state optimized using only the planar energy captures the covariances surprisingly well: the median and largest relative discrepancies from the resummed QPT references at $\lambda=1$ are $2.4\times10^{-3}$ and $3.3\times10^{-3}$, respectively, although no covariance was fitted. Adding one auxiliary coupling reduces the median discrepancy to $1.1\times10^{-4}$; three couplings reduce the largest discrepancy to $2.5\times10^{-4}$. These are differences from resummed estimates, not exact ground-state errors. The residual discrepancies depend on the planar input, the retained fluctuation observables and the resummation, as well as on the auxiliary family.

\begin{table}[tbp]
\centering\small
\begin{tabular}{lrr}
\toprule
Correction to the single-trace state & Median relative difference & Largest relative difference\\
\midrule
None & $2.41\times10^{-3}$ & $3.32\times10^{-3}$\\
One auxiliary coupling & $1.13\times10^{-4}$ & $9.08\times10^{-4}$\\
Three auxiliary couplings & $8.83\times10^{-5}$ & $2.46\times10^{-4}$\\
General Gaussian correction & $7.63\times10^{-5}$ & $1.07\times10^{-4}$\\
\bottomrule
\end{tabular}
\caption{Absolute relative differences from the QPT reference for all ten covariances in \eqref{eqn:two-matrix-connected-covariances}, at $\lambda=1$. Every row uses the same $L=8$, $K=12$ planar state; only the $O(N^0)$ energy is minimized over the correction. No covariance is fitted. For each reference we estimate uncertainty by the larger of half the range of the order-ten approximants and the change in the median from order nine to ten. These relative uncertainty estimates have median $1.8\times10^{-5}$ and maximum $7.4\times10^{-5}$; they are convergence diagnostics, not rigorous bounds.}
\label{tab:two-matrix-covariances}
\end{table}

\begin{figure}[tbp]
\centering
\includegraphics[width=\textwidth]{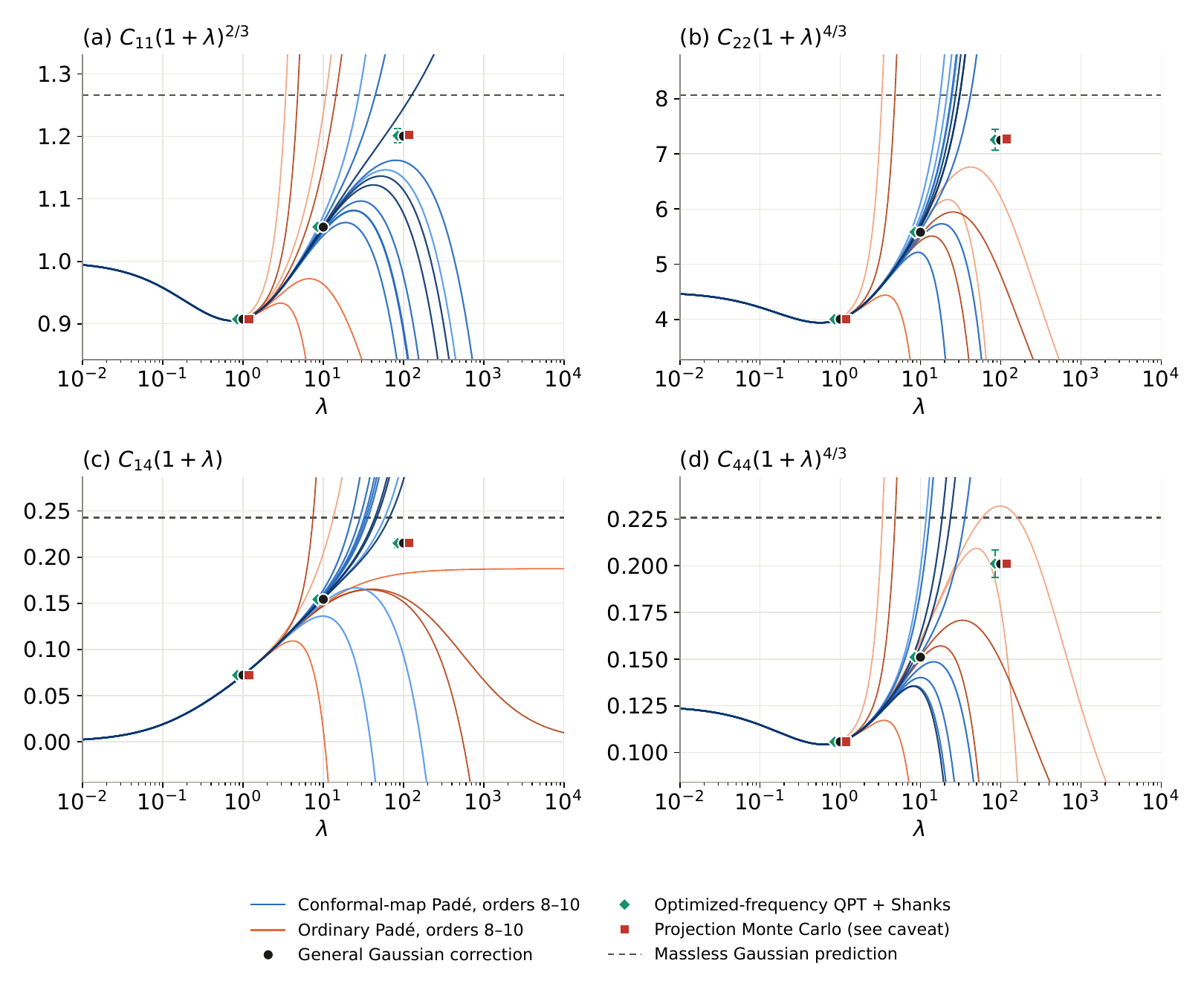}
\caption{Four covariances versus commutator coupling, multiplied by the indicated powers of $1+\lambda$ to approach finite massless values. Orange and blue curves are ordinary and conformally mapped Pad\'e approximants using QPT through orders eight to ten; their variation is not a confidence band. Curves whose Pad\'e denominator vanishes anywhere in the plotted coupling range are excluded. Black circles are the $L=8$, $K=12$ predictions with the general Gaussian correction on $T_1,\ldots,T_4$, at $\lambda=1,10,100$. Green diamonds show optimized-frequency QPT followed by Shanks extrapolation; their bars are the absolute change from the last unextrapolated estimate, not statistical errors. Red squares are the projection Monte Carlo estimates at $\lambda=1,100$; their bars include statistical and extrapolation uncertainties, and at $\lambda=100$ a time-step systematic, but omit cutoff sensitivity and error in the planar SD reference. The green and red points are shifted left and right, respectively, by a factor $1.18$ for visibility. Dashed lines are the separately computed massless predictions with the same cutoffs and Gaussian correction, not exact reference values.}
\label{fig:two-matrix-covariances}
\end{figure}

Figure~\ref{fig:two-matrix-covariances} extends the comparison to stronger coupling. The ordinary and conformally mapped Pad\'e families shown lose stability as the coupling increases. We also reorganize QPT around an adjustable oscillator frequency, choose stationary approximants~\cite{Stevenson:1981Optimized}, and apply Shanks extrapolation~\cite{Shanks:1955Transformations} to the last three available orders. Appendix~\ref{app:resummation} gives the resummation procedures.

\begin{samepage}
The relative differences between the general Gaussian predictions and these estimates are $3\times10^{-4}$--$5\times10^{-4}$ at $\lambda=10$ and $5\times10^{-4}$--$1.0\times10^{-3}$ at $\lambda=100$. However, the estimated relative resummation uncertainties are $3\times10^{-3}$--$1.2\times10^{-2}$ and $9\times10^{-3}$--$3.6\times10^{-2}$, respectively, so this agreement does not establish accuracy at the smaller discrepancy scale. The rescaled massless predictions provide a further comparison, but have no independent precision benchmark here.\par
\end{samepage}

We also compare with ground-state projection Monte Carlo~\cite{Sarsa:2000PathIntegral} at $\lambda=1$ and $100$. The plotted estimates use $N=4,6,8,10$ at $\lambda=1$ and additionally $N=12$ at $\lambda=100$. The trial exponents retain words through $L=6$; the strong-coupling runs also include a quadratic correction to trace fluctuations. To reduce statistical error, we extrapolate the difference between the projected covariance and a reference covariance obtained by reweighting the same paths, then add back the trial covariance computed from the planar SD equations. The bars combine statistical uncertainty and the spread between linear and quadratic fits in $1/N^2$; at $\lambda=100$ they also include an estimated time-step bias from halving the imaginary-time spacing. Appendix~\ref{app:monte-carlo} gives the setup, trajectory diagnostics and normalizability caveats.

These comparisons require a qualification. The polynomial trial states are not globally normalizable, although the stored configuration diagnostics show no sign of escape into the unstable tails. A smooth cutoff outside the sampled region would define normalizable states, but the cutoff dependence of both the projected and reference estimates has not been tested. The quoted bars omit this uncertainty and the systematic error in the planar SD reference. We therefore treat the Monte Carlo results as supporting numerical evidence, without assuming exponentially small corrections from the tails.

At $\lambda=100$, the median relative discrepancy from Monte Carlo across all ten covariances falls from $1.40\times10^{-2}$ for the single-trace state to $2.66\times10^{-3}$ with one auxiliary coupling, $2.17\times10^{-3}$ with three, and $1.85\times10^{-3}$ with the general Gaussian correction. The quoted relative Monte Carlo uncertainties range from $9.3\times10^{-4}$ to $3.73\times10^{-3}$, with median $1.71\times10^{-3}$. These comparisons support the improvement from auxiliaries, but do not establish accuracy at the scale of the remaining discrepancies.

\begin{figure}[p]
\centering
\includegraphics[width=0.97\textwidth]{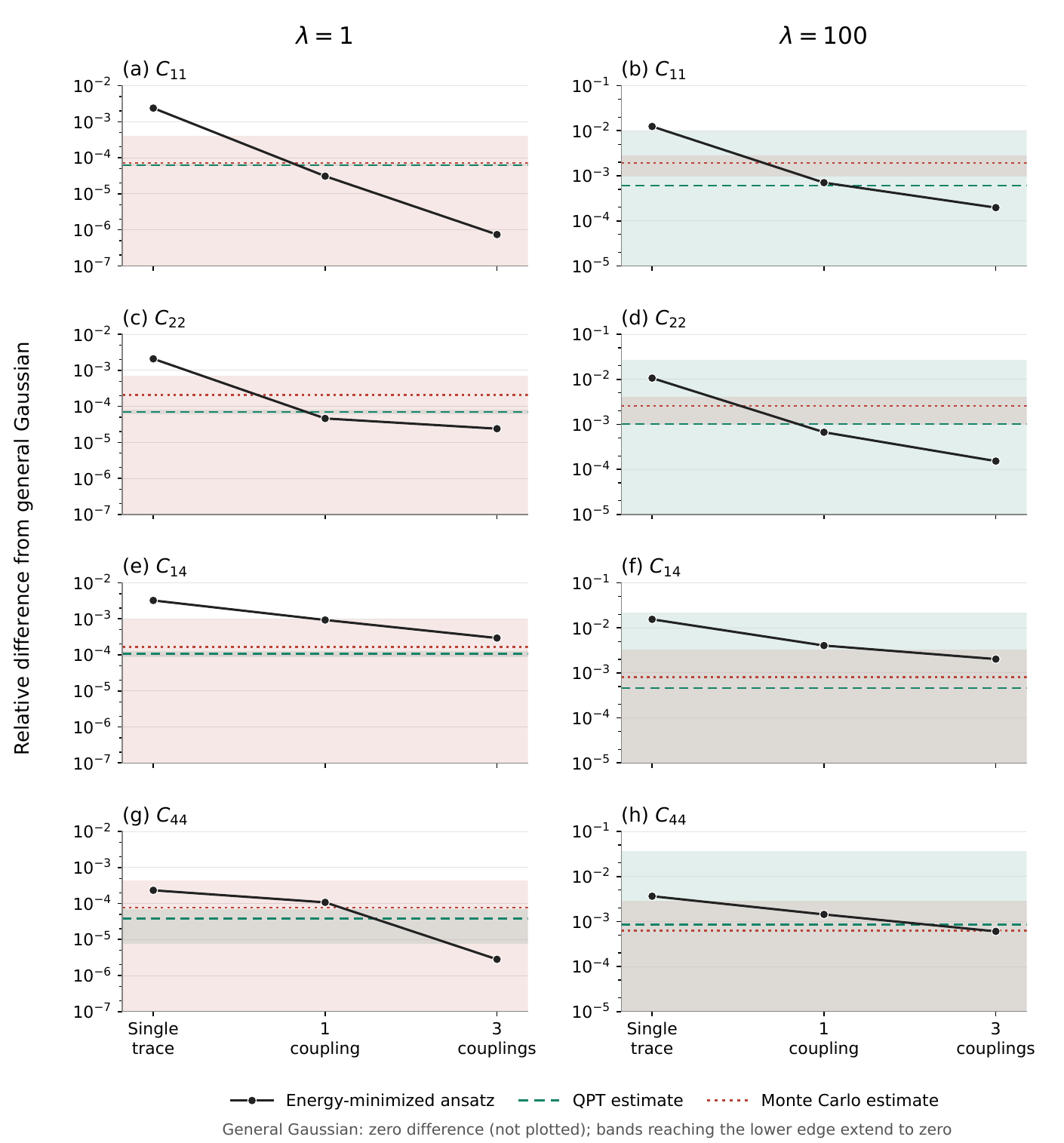}
\caption{Absolute differences divided by the absolute general Gaussian prediction, at $\lambda=1$ (left) and $100$ (right), on logarithmic vertical axes. Black points use the same $L=8$, $K=12$ planar input at each coupling, with the $O(N^0)$ energy minimized over each correction; no covariance is fitted. The reference optimizes all ten real quadratic coefficients on $T_1,\ldots,T_4$. Its zero difference is omitted; it is not an exact ground-state value. Green dashed and red dotted lines compare QPT and Monte Carlo, respectively, with the same reference. Their residual disagreement with the Gaussian prediction likely reflects error in the truncated planar input. QPT uses the conformally mapped Pad\'e estimates of table~\ref{tab:two-matrix-covariances} at $\lambda=1$, and optimized-frequency QPT with Shanks extrapolation at $\lambda=100$ (appendix~\ref{app:resummation}). Shading shows the range of absolute relative differences allowed by each quoted reference interval. Bands reaching the lower edge extend to zero. QPT intervals are resummation diagnostics, not confidence intervals; Monte Carlo intervals omit cutoff sensitivity and error in the planar SD reference (appendix~\ref{app:monte-carlo}).}
\label{fig:two-matrix-covariance-convergence}
\end{figure}

Figure~\ref{fig:two-matrix-covariance-convergence} makes the improvement visible for individual covariances at fixed coupling. One auxiliary coupling captures most of the change in $C_{11}$ and $C_{22}$, while $C_{14}$ changes appreciably as the family is enlarged. The strong-coupling QPT bands are too broad to resolve the later stages of this convergence. The remaining differences between the general Gaussian predictions and QPT or Monte Carlo likely reflect residual error in the fixed $L=8$, $K=12$ planar input, which the Gaussian correction inherits. Increasing the planar cutoffs would test this explanation.

\FloatBarrier

\section{Discussion}

The ans\"atze \eqref{eqn:single-trace} and \eqref{eqn:basic-auxiliary-ansatz}, inspired by putting the `t Hooft expansion front and center, are shockingly effective at capturing the large-$N$ structure of the strongly-coupled MQM ground state. In fact, just the single-trace ansatz \eqref{eqn:single-trace} optimized by minimizing the $O(N^2)$ energy generates connected correlators that are far closer to the true ground state value than they seem to have any right to be. A possible clue is the rapid decrease of the optimized coefficient magnitudes with word length in the normalizations of \eqref{eqn:one-matrix-explicit-polynomial} and \eqref{eqn:two-matrix-explicit-polynomial}. It would be fascinating to understand whether this decrease explains the efficiency of the ansatz, and whether it has a holographic interpretation.

There is a lot of future work to explore. The efficacy of the ans\"atze we consider might also provide some insight into why matrix bootstrap is so effective for capturing the MQM ground state, if the ground state is to good approximation a matrix integral with a simple action. An obvious next step is to extend the technology to the supersymmetric case, to hopefully begin studying the BMN or BFSS ground state directly at large $N$. Another natural next step is computing entanglement entropy and searching for geometry directly in the matrix integral of the optimal variational ansatz. Also, it was recently observed in~\cite{Collier:2026COneMatrixIntegral} that the $c=1$ string, known to be dual to single-matrix MQM~\cite{Das:1990StringField,Gross:1991Fermionic,Ginsparg:1993Lectures}, also admits a perturbative description by a matrix integral; it would be interesting to see if our understanding of the ground state structure as a matrix integral might shed some light on this relationship from a different point of view.

The power of the wavefunctions considered in this paper is also a good motivation for considering them outside of the ground state sector as well. For example, \cite{Fliss:2026Topology} uses auxiliaries to prepare a thermofield-double-like entangled state of two copies of an MQM system; perhaps such wavefunctions would be a better approximation to the thermal density matrix than initially thought. The recent application of bootstrap to BPS sectors containing black hole states~\cite{Chen:2026BPSBootstrap} also suggests that similar ans\"atze might capture BPS black hole physics. Lastly, one interpretation of this work is the existence of a possible close relationship between 0+0 and 0+1 dimensional matrix path integrals. Perhaps this might be generalizable to higher dimensions, with matrix integrals describing to good approximation the emergence of geometry from eigenvalues in the ground state of higher dimensional SYM~\cite{Itzhaki:1998Supergravity}. 

\section*{Acknowledgments}

I am indebted to Zohar Komargodski and Alessio Miscioscia for insightful feedback, and Sean A. Hartnoll for invaluable conversations on this topic over the years. I am funded by the Simons Center for Geometry and Physics. I thank OpenAI for providing complimentary access to frontier models and workspace tools through the ChatGPT for Academic Researchers program. This project was supported by Anthropic. Astra 6 and Opus 5.5 were both used extensively for coding for the numerics (which they have done much faster and more accurately than I could), checking analytic derivations, and manuscript editing -- I take full responsibility for the veracity and understanding of the code. The ideas were fully human.

\appendix

\section{Optimization Algorithms}\label{app:optimization}

\subsection{Single-Trace Optimization}\label{app:single-trace-optimization}

The numerical states in \S3 have real coefficients and the symmetries specified for each model. Write the exponent per $N$ and the action of its squared norm as
\begin{equation}\label{eqn:algorithm-action}
v(X)=N^{-1}f_{\boldsymbol g}(X)=\sum_w g_w w(X),
\qquad S(X)=2v(X).
\end{equation}
Choose the exponent words through length $L$ and the independent cumulants through length $K$. For two matrices, remove cyclic duplicates, identify reversal and $X\leftrightarrow Y$, and discard words odd in either matrix. A fitted coefficient multiplies the sum of the distinct cyclic words related by these symmetries, without dividing by their number. Choose the alphabetically first representative, with $X<Y$, for each cumulant. Writing that representative as $uX^i$, retain the SD equation
\begin{equation}\label{eqn:algorithm-sd}
F_{i,u}=m(u\mathcal D_iS)-\sum_{u=aX^ib}m(a)m(b)=0,
\qquad m(1)=1.
\end{equation}
This gives one equation per cumulant unknown. For $D=9$, contract the same equations with the independent pairings of \S\ref{sec:bosonic-bfss}. Reconstruct every required moment from \eqref{eqn:free-moment-cumulant}, setting only cumulants above $K$ to zero; the SD equations can require moments through length $K+L-2$. Store repeated moment calculations and differentiate the same recursion to obtain $J$ and the coefficient derivatives.

At fixed $\boldsymbol g$, use Newton iteration: solve a linear system with matrix $J$ and right-hand side minus the SD residuals, then add the solution to the cumulants. Stop when the largest absolute residual is below $10^{-13}$ for two matrices and $10^{-12}$ for $D=9$. The initial Gaussian estimate has second cumulant $1/(4g_{X^2})$ and all higher cumulants zero; subsequent evaluations start from the previous solution. Use the optimized exponent coefficients to initialize larger $L$ or $K$, setting newly added exponent coefficients to zero.

For a length-$n$ word $w$ with successive labels $i_1,\ldots,i_n$, define the trace-splitting polynomial
\begin{equation}\label{eqn:algorithm-split}
\begin{gathered}
\omega_w(t)=2\!\sum_{\substack{1\leq r<s\leq n\\i_r=i_s}}
t_{i_{r+1}\cdots i_{s-1}}\,
t_{i_{s+1}\cdots i_n i_1\cdots i_{r-1}},
\qquad t_{\varnothing}=1,\\
\omega_v(t)=\sum_w g_w\,\omega_w(t).
\end{gathered}
\end{equation}
Here $t_w$ is a normalized trace moment treated as a variable, and $v$ is the entire exponent polynomial in \eqref{eqn:algorithm-action}, not an individual word. We extend $\omega$ linearly to any matrix polynomial, as shown for $v$ in the second line. Substituting the computed moments, $t_w=m(w)$, gives the objective used in the numerical minimization:
\begin{equation}\label{eqn:algorithm-energy}
\begin{aligned}
\mathcal E_{\rm SD}(\boldsymbol\kappa,\boldsymbol q)&=m(V)+\frac14\omega_v(m),\\
e_{L,K}(\boldsymbol q)&=\mathcal E_{\rm SD}(\boldsymbol\kappa(\boldsymbol q),\boldsymbol q).
\end{aligned}
\end{equation}
Here the moments are reconstructed from $\boldsymbol\kappa$. Integration by parts gives equality with the kinetic-moment expression in \eqref{eqn:variational-energy} when the required SD identities hold. A finite selection of SD equations need not preserve this equality; we use \eqref{eqn:algorithm-energy} consistently for both the objective and its gradient, computed by \eqref{eqn:sd-energy-gradient}. For two matrices we use the Broyden--Fletcher--Goldfarb--Shanno (BFGS) method~\cite[\S6.1]{Nocedal:2006NumericalOptimization}, which estimates the inverse matrix of second derivatives from successive gradients, initially using half the identity matrix. Halve a proposed step until the SD solve succeeds and the energy drop is at least $10^{-4}$ times that predicted by the gradient. The target largest gradient component is $10^{-11}$. Up to four final Newton steps estimate second derivatives by central differences: vary each coefficient by $+10^{-5}$ and $-10^{-5}$, subtract the resulting gradients, and divide by $2\times10^{-5}$. The $D=9$ runs use Newton steps directly, with step reduction and target $10^{-10}$; the $K=10$ run reuses the second-derivative matrix from $K=8$. Small gradients do not establish convergence with the cutoffs or global normalizability.

The restricted length-ten reference in \eqref{eqn:two-matrix-internal-reference} uses seven polynomials, with one fitted coefficient for each:
\begin{equation}\label{eqn:reference-seven-polynomials}
\begin{gathered}
v_{\rm ref}=\sum_{j=1}^{7}a_jP_j,\\
P_1=X^2+Y^2,\qquad P_2=X^4+Y^4,\qquad
P_3=X^2Y^2,\qquad P_4=XYXY.
\end{gathered}
\end{equation}
The first four span the symmetry-allowed terms through length four. The remaining three are chosen from the products of cyclic derivatives generated when the kinetic term acts on the wavefunction. Let $v_d$ denote the part of the current optimized exponent containing words of length $d$, and write
\begin{equation}\label{eqn:reference-quadratic-part}
v_2=\frac{\alpha}{2}(X^2+Y^2),\qquad \alpha=2g_{X^2}.
\end{equation}
Cancelling the next-degree products against the cross term between a new polynomial and $v_2$ suggests
\begin{equation}\label{eqn:reference-polynomial-construction}
\begin{aligned}
P_5&=-\frac{1}{12\alpha}\sum_{i=X,Y}(\mathcal D_i v_4)^2,\\
P_6&=-\frac{1}{8\alpha}\sum_{i=X,Y}(\mathcal D_i v_4)(\mathcal D_i v_6),\\
P_7&=-\frac{1}{10\alpha}\sum_{i=X,Y}\left[(\mathcal D_i v_4)(\mathcal D_i v_8)+\frac12(\mathcal D_i v_6)^2\right].
\end{aligned}
\end{equation}
These identities are understood inside a trace. Each right-hand side is evaluated when that polynomial is first added, using the optimized states through length four, six and eight, respectively, at $K=12$. We then hold the polynomial fixed and refit all retained coefficients before constructing the next one. This cancellation selects one combination at each new length; trace splitting from still longer exponent terms is omitted when choosing it, and the energy \eqref{eqn:algorithm-energy} determines its fitted coefficient.

Using the symmetry sum $\mathcal S$ of \eqref{eqn:two-matrix-word-sum}, extended linearly to polynomials, the resulting fixed combinations are as follows. All displayed coefficients are rounded.
\begingroup
\small
\begin{equation}\label{eqn:reference-p5}
\begin{aligned}
P_5=10^{-2}\mathcal S\bigl[&-0.000237488327346 X^{6} -0.403378482002 X^{4}Y^{2}\\
&+1.72763830225 X^{3}YXY -1.32497228459 X^{2}YX^{2}Y\bigr].
\end{aligned}
\end{equation}
\begin{equation}\label{eqn:reference-p6}
\begin{aligned}
P_6=10^{-3}\mathcal S\bigl[&-2.87863991231\times10^{-6} X^{8} +0.128651333543 X^{6}Y^{2}\\
&-0.817495253196 X^{5}YXY +2.11719327681 X^{4}YX^{2}Y\\
&+0.259633616448 X^{4}Y^{4} -1.42836087168 X^{3}YX^{3}Y\\
&-1.38349372863 X^{3}YXY^{3} +1.68752763058 X^{3}Y^{2}XY^{2}\\
&-3.12822511966 X^{2}YXYXY^{2} +3.99290696569 X^{2}YXY^{2}XY\\
&+0.262893233208 X^{2}Y^{2}X^{2}Y^{2} +1.13293134830 (XY)^4\bigr].
\end{aligned}
\end{equation}
\begin{equation}\label{eqn:reference-p7}
\begin{aligned}
P_7=10^{-4}\mathcal S\bigl[&-1.29968484691\times10^{-7} X^{10} -0.0430549681033 X^{8}Y^{2}\\
&+0.365883981673 X^{7}YXY -1.33854406930 X^{6}YX^{2}Y\\
&-0.130747725209 X^{6}Y^{4} +2.75044517151 X^{5}YX^{3}Y\\
&+0.834231736003 X^{5}YXY^{3} -0.843633839215 X^{5}Y^{2}XY^{2}\\
&-1.73473076562 X^{4}YX^{4}Y -1.72453896497 X^{4}YX^{2}Y^{3}\\
&+2.51078634155 X^{4}YXYXY^{2} -3.38823072884 X^{4}YXY^{2}XY\\
&-0.308368869506 X^{4}Y^{2}X^{2}Y^{2} +2.32188395883 X^{3}YX^{3}Y^{3}\\
&-4.82035443579 X^{3}YX^{2}YXY^{2} +4.82223358541 X^{3}YX^{2}Y^{2}XY\\
&+2.40766517393 X^{3}YXYX^{2}Y^{2} -5.45745351811 X^{3}YXYXYXY\\
&-0.574572269598 X^{3}Y^{2}X^{3}Y^{2} -3.81874268320 X^{2}YX^{2}YX^{2}Y^{2}\\
&+7.70708405350 X^{2}YX^{2}YXYXY -3.56726655059 X^{2}YXYX^{2}YXY\bigr].
\end{aligned}
\end{equation}
\endgroup
The seven fitted multipliers at $\mu=1$ and $K=12$ are
\begin{equation}\label{eqn:reference-seven-coefficients}
\begin{aligned}
a_1&=0.583855563955,&\qquad a_2&=-0.000686802797502,\\
a_3&=0.197703768294,& a_4&=-0.199077368899,\\
a_5&=1.11337493481,& a_6&=1.12065923141,\\
a_7&=0.685098552960.
\end{aligned}
\end{equation}

The one-matrix planar results instead use the density \eqref{eqn:one-matrix-trial-density}. Expand $Q_{L-2}(a\cos\theta)$ in $\cos(2r\theta)$ for $r=0,\ldots,L/2-1$. At fixed endpoint $a$, minimize \eqref{eqn:one-matrix-reference-observables} over these coefficients by Newton iteration with the linear constraint $\int\rho_L=1$; then vary $a$ until the derivative of the minimized energy vanishes. We use midpoint quadrature: divide $0\leq\theta\leq\pi$ into equal intervals and approximate the integral by the sum of their widths times the integrand at each midpoint. We use $2L+8$ intervals for the energy and $2L+64$ for the moments through $X^{40}$. Only nonnegative densities with normalizable exponents are reported.

\subsection{Auxiliary Optimization}\label{app:auxiliary-optimization}

For two matrices, fix the optimized planar exponent. Supplement the four observables $T_a$ in \eqref{eqn:two-matrix-connected-covariances} with all symmetry-allowed trace sums through length $L+4-2$ (length ten for $L=8$). Denote the corresponding matrix polynomials and trace observables by
\begin{equation}\label{eqn:algorithm-observables}
U_\alpha=\Tr p_\alpha,\qquad t_\alpha=m(p_\alpha).
\end{equation}
The first four $U_\alpha$ are the $T_a$. We first compute the leading covariance matrix $\widehat C_0$ of this enlarged list in the single-trace state, before adding auxiliaries.

To measure how these moments respond to changes in the matrix-integral weight, introduce a real parameter $\xi_\beta$ for each polynomial $p_\beta$:
\begin{equation}\label{eqn:algorithm-sources}
S_{\boldsymbol\xi}=S+\sum_\beta\xi_\beta p_\beta.
\end{equation}
These parameters are called sources. The modified weight is $e^{-N\Tr S_{\boldsymbol\xi}}$, and all source derivatives below are evaluated at $\boldsymbol\xi=0$, where it reduces to the original weight. Since $S=2v$, a source shifts the coefficient of $p_\beta$ in the wavefunction exponent $v$ by $\xi_\beta/2$. We hold the optimized coefficients fixed while taking this derivative; we do not minimize the energy again.

For a simple example, take one matrix and add $\xi X^2$ to $S$. At finite $N$, define the normalized second moment by
\begin{equation}\label{eqn:algorithm-m2-source}
m_{2,N}(\xi)=\frac{\displaystyle\int dX\,\frac1N\Tr X^2\,e^{-N\Tr(S+\xi X^2)}}{\displaystyle\int dX\,e^{-N\Tr(S+\xi X^2)}}.
\end{equation}
Differentiating both numerator and denominator gives
\begin{equation}\label{eqn:algorithm-m2-covariance}
-\frac{d m_{2,N}}{d\xi}
=\left\langle(\Tr X^2)^2\right\rangle_\xi
-\left\langle\Tr X^2\right\rangle_\xi^2.
\end{equation}
Here the brackets denote normalized expectations in the modified weight. Differentiating the exponential inserts $-N\Tr X^2$; the factor $N$ cancels the $1/N$ in the moment. Differentiating the denominator produces the subtraction of the disconnected product. At $\xi=0$, the right-hand side is therefore the variance of $\Tr X^2$ in the original state. This identity holds for the convergent integral; taking its smooth planar limit relates the derivative of the planar moment $m_2$ to the leading $O(N^0)$ variance.

The same argument with $p_\alpha$ and $p_\beta$ gives
\begin{equation}\label{eqn:algorithm-response}
\begin{aligned}
m_{\boldsymbol\xi}(p_\alpha)&=\lim_{N\to\infty}\frac1N\langle U_\alpha\rangle_{\boldsymbol\xi},\\
(\widehat C_0)_{\alpha\beta}
&=\lim_{N\to\infty}\left(\langle U_\alpha U_\beta\rangle_0-\langle U_\alpha\rangle_0\langle U_\beta\rangle_0\right)
=-\left.\frac{\partial m_{\boldsymbol\xi}(p_\alpha)}{\partial\xi_\beta}\right|_{\boldsymbol\xi=0}.
\end{aligned}
\end{equation}
Thus a covariance can be extracted from the response of a planar one-trace expectation value. For the numerical states treated as local planar solutions, we compute this response on the chosen branch, with the normalizability qualifications of \S2.2.

Numerically, we obtain the derivative by differentiating the SD equations. Let $\boldsymbol F_K$ collect the residuals in \eqref{eqn:algorithm-sd}, with every moment reconstructed from the retained cumulants. The source-dependent solution obeys
\begin{equation}\label{eqn:algorithm-source-sd}
\boldsymbol F_K(\boldsymbol\kappa(\boldsymbol\xi),\boldsymbol\xi)=0,
\qquad J=\frac{\partial\boldsymbol F_K}{\partial\boldsymbol\kappa}.
\end{equation}
Applying the chain rule gives a linear system for the change of each cumulant:
\begin{equation}\label{eqn:algorithm-source-linear-system}
J\,\frac{\partial\boldsymbol\kappa}{\partial\xi_\beta}
=-\left.\frac{\partial\boldsymbol F_K}{\partial\xi_\beta}\right|_{\boldsymbol\kappa\,\mathrm{fixed}},
\qquad
\left.\frac{\partial F_{i,u}}{\partial\xi_\beta}\right|_{\boldsymbol\kappa\,\mathrm{fixed}}
=m_{\boldsymbol\kappa}(u\mathcal D_i p_\beta).
\end{equation}
All quantities in this equation are evaluated at the original solution with $\boldsymbol\xi=0$. The derivative on the right acts only on the explicit source in $S_{\boldsymbol\xi}$, holding the cumulants fixed. We solve this system once for each source, reusing the matrix $J$ from the planar SD solve. Differentiating the moment--cumulant relation then gives
\begin{equation}\label{eqn:algorithm-source-chain-rule}
(\widehat C_0)_{\alpha\beta}
=-\sum_w\frac{\partial m_{\boldsymbol\kappa}(p_\alpha)}{\partial\kappa_w}
\frac{\partial\kappa_w}{\partial\xi_\beta}.
\end{equation}
The sum runs over the independent cumulants retained through length $K$, and the moment derivatives are obtained by differentiating the same polynomials used to reconstruct the moments. No solutions at nearby nonzero sources are required: the derivative follows from a linear system at the original solution. Finite differences of nearby SD solutions could provide an independent check, but are not the derivative prescription used here. At finite $K$, this procedure gives the covariance estimate of the truncated SD system. Average the numerical covariance matrix with its transpose; its first $4\times4$ submatrix is $C_0$ in \eqref{eqn:auxiliary-covariance}.

To construct $R$ in \eqref{eqn:auxiliary-energy}, form the SD residuals and moment derivatives
\begin{equation}\label{eqn:algorithm-local-polynomials}
\begin{aligned}
\sigma_a(t)&=\frac12\omega_{p_a}(t)-\sum_i t\!\left((\mathcal D_i v)(\mathcal D_i p_a)\right),\qquad a=1,\ldots,4,\\
G_{a\alpha}&=\left.\frac{\partial\sigma_a}{\partial t_\alpha}\right|_{t=m},\qquad
\Lambda_{\alpha\beta}=\left.\frac12\frac{\partial^2\omega_v}{\partial t_\alpha\partial t_\beta}\right|_{t=m}.
\end{aligned}
\end{equation}
Here $t(p)$ denotes the linear combination of moments in $p$. Express individual word moments in the chosen symmetry sums; for example, the moment of $X^2$ is half that of $p_1$. Treat the $t_\alpha$ as independent coordinates when differentiating, holding the exponent coefficients fixed; only afterward evaluate them at the optimized moments in \eqref{eqn:algorithm-observables}. Thus these derivatives are taken before substituting the moment--cumulant relation. The matrix $\Lambda$ comes from the trace-splitting term obtained by applying $H$ to the state and dividing by the state; the other terms are linear in moments and give no second derivatives. Define
\begin{equation}\label{eqn:algorithm-R}
\begin{aligned}
P_0&=C_0^{-1},\qquad \Gamma_{\alpha a}=\sum_{b=1}^4(\widehat C_0)_{\alpha b}(P_0)_{ba},\\
R&=-\frac14P_0\Omega P_0-\frac12\left(P_0G\Gamma+\Gamma^{\mathsf T}G^{\mathsf T}P_0\right)
+\Gamma^{\mathsf T}\Lambda\Gamma.
\end{aligned}
\end{equation}
Here $\Omega$ is defined in \S2.2. In the Gaussian fluctuation distribution of the single-trace state, $\Gamma$ determines the average of the full set of centered fluctuations when the four retained fluctuations are held fixed:
\begin{equation}\label{eqn:algorithm-conditional-mean}
\delta U_\alpha=U_\alpha-Nm(p_\alpha),\qquad
\mathbb E_0[\delta U_\alpha\mid\delta\boldsymbol T]
=\sum_{a=1}^{4}\Gamma_{\alpha a}\delta T_a.
\end{equation}
The conditional expectation on the left averages over the remaining fluctuations at fixed $\delta\boldsymbol T$. Their conditional covariance is independent of those fixed values and cancels from the energy difference. Equation~\eqref{eqn:algorithm-R} therefore includes the effect of the unvaried observables without further matrix integrals.

For either short auxiliary family in \eqref{eqn:two-matrix-short-auxiliaries}, expand
\begin{equation}\label{eqn:algorithm-logdet}
\frac12\log\det\mathcal M_{\boldsymbol\eta}
=\frac12\sum_{n\geq1}\frac{(-1)^{n+1}}{n}
\operatorname{Tr}_{A}\!\left[(\mathcal M_{\boldsymbol\eta}-I)^n\right].
\end{equation}
Here $I$ is the identity map and $\operatorname{Tr}_A$ traces over matrix components: a term mapping $A$ to $a(X)A b(X)$ contributes $\Tr a\,\Tr b$. Keep products with both traces symmetry allowed and of length at most four; terms beyond $n=4$ cannot contribute. Center the traces, compensate linear terms, and extract $B_{\boldsymbol\eta}$ using \eqref{eqn:auxiliary-covariance}. This finite polynomial prescription does not assume convergence of the full logarithmic series.

Minimize \eqref{eqn:auxiliary-energy}, rejecting couplings for which $P_0+2B_{\boldsymbol\eta}$ is not positive definite. We use Nelder--Mead, which compares energies at nearby parameter points without derivatives, starting from every point of $\{-0.1,-0.02,0.02,0.1\}^d$, where $d=1$ or $3$ is the number of auxiliary couplings, and keep the lowest-energy result.\footnote{The implementation minimizes $10^6\Delta E_1$, with parameter and objective tolerances $10^{-10}$ and $10^{-14}$, and at most 4000 iterations per start. These are stopping settings, not uncertainty estimates.} For the unrestricted real Gaussian correction, the positive solution is
\begin{equation}\label{eqn:algorithm-gaussian-minimum}
P_*=2\Omega^{-1/2}\left(\Omega^{1/2}R\Omega^{1/2}\right)^{1/2}\Omega^{-1/2},
\qquad C_*=P_*^{-1},\qquad B_*=\frac12(P_*-P_0).
\end{equation}
The square roots are the positive matrix square roots; this formula requires positive $\Omega$ and $R$. Covariances are then predictions of the energy-minimized state, with the cutoff and normalizability qualifications of \S2.2.

For the finite-$N$ one-matrix test, we instead optimize all coefficients of $s$ and $u$ in \eqref{eqn:one-matrix-auxiliary-polynomials} together. Orthonormalize the $N$ functions $\sqrt{u'(x)}\,u(x)^k e^{-Ns(x)}$, $k=0,\ldots,N-1$, and let $P$ project onto their span. Then
\begin{equation}\label{eqn:algorithm-fermions}
h=-\frac{1}{2N}\frac{d^2}{dx^2}+NV(x),\qquad
E=\operatorname{tr}(Ph),\qquad C_{jk}=\operatorname{tr}\!\left(Px^j(1-P)x^k\right).
\end{equation}
Here the traces act on one-particle functions. Build the orbital basis by successive multiplication by $u$. At each step, subtract the projections onto all earlier normalized orbitals twice, then normalize the new function. Repeating the subtraction reduces numerical loss of orthogonality. The optimization uses BFGS followed by Nelder--Mead on a uniform grid with $\max(1200,60N)$ points in $[-2.6,2.6]$. For an orbital $\phi$, grid points $x_j$ and spacing $\Delta x$, the second derivative is approximated by
\begin{equation}\label{eqn:algorithm-grid-derivative}
\phi''(x_j)\simeq
\frac{-\phi(x_{j+2})+16\phi(x_{j+1})-30\phi(x_j)+16\phi(x_{j-1})-\phi(x_{j-2})}{12(\Delta x)^2}.
\end{equation}
This five-point formula has error $O((\Delta x)^4)$ in the interior; values required outside the interval are set to zero.

The reported states are checked for $u'(x)>0$ on the whole real axis and reevaluated by direct integration with analytic orbital derivatives. An independent reference diagonalizes $h$ in the first 128, 192 and 256 harmonic-oscillator states of frequency $2.5$, with the kinetic term in \eqref{eqn:algorithm-fermions}. The 256-state result is used for comparison; increasing from 192 to 256 changes the reported reference covariances by less than $2\times10^{-14}$ in relative terms. This test uses the full finite-$N$ energy, without SD or a Gaussian approximation to the trace fluctuations.
\begingroup
\interfootnotelinepenalty=10000
\clubpenalty=10000
\widowpenalty=10000
\section{Quantum Perturbation Theory and Resummation}\label{app:resummation}

The independent two-matrix references in \S\ref{sec:two-matrix-covariances} start from the weak-coupling expansions of the covariances in \eqref{eqn:two-matrix-connected-covariances},
\begin{equation}\label{eqn:qpt-covariance-series}
C_{ab}(\lambda)=\sum_{k=0}^{10}c_{ab,k}\lambda^k+O(\lambda^{11}).
\end{equation}
We expand the logarithm of the ground-state wavefunction and its energy:
\begin{equation}\label{eqn:qpt-wavefunction-expansion}
\begin{gathered}
\Psi_\lambda=e^{-W_\lambda},\qquad
W_\lambda=\sum_{n\geq0}\lambda^nW_n,\qquad
E_\lambda=\sum_{n\geq0}\lambda^nE_n,\\
W_0=\frac N2\Tr(X^2+Y^2).
\end{gathered}
\end{equation}
Here $\nabla$ and $\Delta$ are the gradient and Laplacian with respect to the independent real matrix components, in a traceless Hermitian basis orthonormal under the trace inner product. Substitution into the Schr\"odinger equation gives, for $n\geq1$,
\begin{equation}\label{eqn:qpt-wavefunction-recursion}
\left(\frac{\Delta}{2N}-\frac{\nabla W_0}{N}\cdot\nabla\right)W_n
=E_n+\frac{N\delta_{n1}}2\Tr[X,Y]^2
+\frac1{2N}\sum_{k=1}^{n-1}\nabla W_k\cdot\nabla W_{n-k}.
\end{equation}
The interaction term is present only at $n=1$. At leading order in $N$, each $W_n$ is $N^2$ times a polynomial in normalized traces; we keep all products generated by this recursion. The term containing $\nabla W_0$ multiplies a degree-$d$ monomial by $-d$, while $\Delta$ lowers its degree by two. Thus $W_n$ is determined from degree $2n+2$ downward, and the constant part of the equation fixes $E_n$. Additive constants in $W_n$ only fix normalization and are set to zero.

We then solve the SD equations for $|\Psi_\lambda|^2$ order by order in $\lambda$. In each product of traces, the matrix derivative acts on each trace in turn; at large $N$, the other normalized traces become planar moments. The source derivatives of \S\ref{app:auxiliary-optimization} give the covariances, holding the coefficients of $W_\lambda$ fixed while including the source dependence of these moment factors. This calculation uses neither the variational coefficients nor the cutoffs $L$ and $K$. As a normalization check,
\begin{equation}\label{eqn:qpt-covariance-check}
C_{11}(\lambda)=1-\frac54\lambda+\frac{199}{64}\lambda^2+O(\lambda^3).
\end{equation}

\subsection{Pad\'e Resummation}\label{app:pade}

A Pad\'e approximant replaces the truncated series by a ratio of polynomials of degrees at most $p$ and $q$, determined by matching its expansion through order $p+q$~\cite{Baker:1996Pade}:
\begin{equation}\label{eqn:qpt-pade}
[p/q](\lambda)=\frac{P_p(\lambda)}{Q_q(\lambda)},\qquad
Q_q(0)=1,\qquad
Q_q(\lambda)C_{ab}(\lambda)-P_p(\lambda)=O(\lambda^{p+q+1}).
\end{equation}
The ordinary approximants in figure~\ref{fig:two-matrix-covariances} use total orders eight, nine and ten, with $q\geq1$ and $|p-q|\leq2$. For the mapped approximants, first substitute the inverse of \eqref{eqn:covariance-pade-map},
\begin{equation}\label{eqn:qpt-inverse-map}
\lambda=\frac{4\rho w}{(1-w)^2},\qquad \rho\in\{0.19,0.21,0.23\},
\end{equation}
reexpand through the same order in $w$, and construct $[p/q]$ in $w$ with $|p-q|\leq1$. The variation with $\rho$ tests sensitivity to this empirical choice of map. Both procedures act directly on the coupling series.

At $\lambda=1$, table~\ref{tab:two-matrix-covariances} uses the median of the three mapped $[5/5]$ values after excluding approximants with nearby poles.\footnote{For each $\rho$, reject a denominator root $z$ with $-0.05\leq\operatorname{Re}z\leq w(1)+0.15$ and $|\operatorname{Im}z|<0.15$, unless a numerator root lies within $10^{-3}\max(1,|z|)$. Failed fits or root searches are also excluded. This selection applies to the tabulated reference. For figure~\ref{fig:two-matrix-covariances}, we omit an entire curve if its denominator has a real zero in the displayed interval $10^{-2}\leq\lambda\leq10^4$, even if the numerator has a nearby zero. For mapped approximants, this test uses the corresponding real $w$ interval.} We estimate uncertainty by the larger of half their range and the change from the median of the accepted order-nine $[4/5]$ and $[5/4]$ values at the same three $\rho$. The tabulated Pad\'e fits use 40-digit arithmetic on the stored double-precision series coefficients. Neither this spread nor the variation of the plotted curves is a rigorous error bound. As an MQM precedent, Asano et al.~\cite{Asano:2018BMNPhase} applied Pad\'e resummation to the three-loop BMN transition-temperature series, with an assumed small-mass scaling; their observable and resummation prescription differ from ours.

\subsection{Optimized Frequency}\label{app:optimized-frequency}

An alternative is to reorganize the same series around an adjustable oscillator frequency. Related MQM calculations use Gaussian actions fixed by gap equations, which impose self-consistency on their two-point functions~\cite{Kabat:2000Approximations}, or perturbation theory in a basis of oscillator states with an optimized frequency~\cite{Hynek:2016Optimized}. Here we choose the frequency by requiring that the truncated result be stationary under its variation, the principle of minimal sensitivity~\cite{Stevenson:1981Optimized}.

Temporarily multiply the quadratic potential in \eqref{eqn:two-matrix-covariance-model} by $m^2$, with $m>0$. Rescaling the matrices gives
\begin{equation}\label{eqn:qpt-mass-scaling}
\begin{gathered}
C_{ab}(m,\lambda)=m^{-s_{ab}}C_{ab}(1,\lambda/m^3),\\
s_{ab}=\frac{d_a+d_b}{2},\qquad (d_1,d_2,d_3,d_4)=(2,4,4,4),
\end{gathered}
\end{equation}
where $d_a$ is the degree of $T_a$. Introduce a trial frequency $\omega>0$ and a formal expansion parameter $\delta$ through
\begin{equation}\label{eqn:qpt-frequency-interpolation}
m^2=\omega^2+\delta(1-\omega^2),\qquad \lambda\longmapsto\delta\lambda.
\end{equation}
At $\delta=0$ the theory is Gaussian with frequency $\omega$; at $\delta=1$ it is the original model. Expand through order $\delta^n$, then set $\delta=1$. This gives
\begin{equation}\label{eqn:qpt-frequency-series}
C_{ab}^{(n)}(\lambda;\omega)=
\sum_{k=0}^{n}c_{ab,k}\lambda^k\omega^{-s_{ab}-3k}
\sum_{j=0}^{n-k}\binom{-(s_{ab}+3k)/2}{j}
\left(\frac{1-\omega^2}{\omega^2}\right)^j.
\end{equation}
The generalized binomial coefficient is defined, for real $\alpha$ and integer $j\geq1$, by
\begin{equation}\label{eqn:qpt-generalized-binomial}
\binom{\alpha}{0}=1,\qquad
\binom{\alpha}{j}=\frac{\alpha(\alpha-1)\cdots(\alpha-j+1)}{j!}.
\end{equation}
For each covariance, coupling and order separately, solve
\begin{equation}\label{eqn:qpt-stationary-frequency}
\partial_\omega C_{ab}^{(n)}(\lambda;\omega)=0.
\end{equation}
We search for sign changes on 240 logarithmically spaced frequencies between $0.2$ and $6\max(1,\lambda)^{1/3}$, refine the roots using 40-digit arithmetic, and choose the root with the smallest absolute second derivative with respect to $\omega$. If none is found, that order is omitted. Thus the frequency is chosen for each observable, rather than by minimizing an energy.

\begin{samepage}
Finally, apply the Shanks transformation\footnote{The author thanks ChatGPT for suggesting this trick, which works surprisingly well.}~\cite{Shanks:1955Transformations} to the last three available stationary estimates, denoted $Q_1,Q_2,Q_3$ in increasing perturbative order:
\begin{equation}\label{eqn:qpt-shanks}
Q_{\mathrm S}=Q_3-\frac{(Q_3-Q_2)^2}{Q_3-2Q_2+Q_1},\qquad
\Delta Q=|Q_{\mathrm S}-Q_3|.
\end{equation}
For the points in figure~\ref{fig:two-matrix-covariances}, these are orders five, seven and nine for $C_{11},C_{22},C_{44}$, and six, eight and ten for $C_{14}$.\footnote{The plotted Shanks values use the archived stationary estimates, stored to six significant digits.} The bars show $\Delta Q$. A small denominator can make the extrapolation unstable, and its change from the last estimate is only a convergence diagnostic, not a statistical uncertainty or a proven bound.
\end{samepage}
\endgroup
\begingroup
\interfootnotelinepenalty=10000
\clubpenalty=10000
\widowpenalty=10000
\section{Monte Carlo Setup}\label{app:monte-carlo}

\subsection{Projection and Measurements}
We use open imaginary-time paths to project a trial wavefunction toward the ground state~\cite{Sarsa:2000PathIntegral}. The matrices are traceless Hermitian, and all simulation parameters below use the unit-quartic-coupling variables of \eqref{eqn:two-matrix-covariance-scaling}:
\begin{equation}\label{eqn:mc-hamiltonian}
\begin{gathered}
Q=(\widetilde X,\widetilde Y),\qquad
\widetilde H=\lambda^{-1/3}H_\lambda=-\frac{\Delta}{2N}+U(Q),\\
U(Q)=N\Tr\!\left[\frac\mu2(\widetilde X^2+\widetilde Y^2)-\frac12[\widetilde X,\widetilde Y]^2\right].
\end{gathered}
\end{equation}
Here $\Delta$ is the flat Laplacian on the independent traceless matrix components. We write the real trial wavefunction in the form
\begin{equation}\label{eqn:mc-trial}
\Psi_T(Q)=e^{-W(Q)},\qquad
W=\Tr f_{\boldsymbol g}(Q)+\frac12\delta\boldsymbol T^{\mathsf T}B\delta\boldsymbol T.
\end{equation}
The four trace fluctuations are those of \eqref{eqn:two-matrix-connected-covariances}, evaluated in the rescaled variables and centered on their planar means. Both couplings use $L=6$, $K=12$: at $\lambda=1$ we set $B=0$, while at $\lambda=100$ we use the general Gaussian correction. These trial states differ from the $L=8$ states used for the variational comparisons.

For a normalizable trial state with nonzero ground-state overlap, midpoint measurements approach the ground-state expectation values as the total projection time increases. We discretize imaginary time into $M+1$ equally spaced points, called imaginary-time slices. Each slice contains a pair of complete $N\times N$ matrices: \begin{equation}\label{eqn:mc-slices} \tau_j=j\epsilon,\qquad Q_j=\bigl(\widetilde X(\tau_j),\widetilde Y(\tau_j)\bigr),\qquad j=0,\ldots,M,\qquad \beta=M\epsilon. \end{equation} Here $\epsilon$ is the imaginary-time spacing and $\beta$ is the total projection time; each sampled path contains all $M+1$ slices, including both endpoints. The discretized weight is $e^{-S_\epsilon}$, where
\begin{equation}\label{eqn:mc-path-action}
S_\epsilon=W(Q_0)+W(Q_M)
+\frac{N}{2\epsilon}\sum_{j=0}^{M-1}\|Q_{j+1}-Q_j\|^2
+\epsilon\sum_{j=0}^{M}\alpha_j U(Q_j).
\end{equation}
The squared norm is the sum of the traces of the squared matrix differences; $\alpha_0=\alpha_M=1/2$, with all other weights equal to one. This symmetric discretization has leading $O(\epsilon^2)$ error under the usual convergence assumptions. Our polynomial trial states do not satisfy global normalizability, so this is a formal weight; \S\ref{app:mc-normalizability} describes the resulting limitation.

We update the entire path by hybrid Monte Carlo (HMC)~\cite{Duane:1987HybridMonteCarlo}; ref.~\cite{Neal:2011HMC} gives a pedagogical account. Let $q$ collect the real coefficients of all matrices at all slices in a traceless Hermitian basis orthonormal under the trace inner product. At each update we draw a zero-mean Gaussian vector $p$ with covariance $\mathsf A$. These auxiliary sampling variables are distinct from the physical MQM momenta. Their Hamiltonian is \begin{equation}\label{eqn:mc-hmc-energy} \mathcal H_{\rm HMC}(q,p)=S_\epsilon(q)+\frac12p^{\mathsf T}\mathsf A^{-1}p. \end{equation} We choose $\mathsf A$ as the Hessian, or matrix of second derivatives with respect to $q$, of a Gaussian path action obtained from \eqref{eqn:mc-path-action} by replacing $W,U$ with \begin{equation}\label{eqn:mc-gaussian-action} \begin{gathered} W_G(Q)=Ng_2\Tr(\widetilde X^2+\widetilde Y^2),\qquad \omega=2g_2,\\ U_G(Q)=\frac{N\omega^2}{2}\Tr(\widetilde X^2+\widetilde Y^2). \end{gathered} \end{equation} Here $g_2$ is the common quadratic coefficient in the single-trace exponent; $\mathsf A$ is constant throughout the run. One leapfrog step is a half-step for $p$, a full step for $q$, and another half-step for $p$: \begin{equation}\label{eqn:mc-leapfrog} \begin{aligned} p_{1/2}&=p-\frac{\eta}{2}\nabla_q S_\epsilon(q),\\ q'&=q+\eta\mathsf A^{-1}p_{1/2},\\ p'&=p_{1/2}-\frac{\eta}{2}\nabla_q S_\epsilon(q'). \end{aligned} \end{equation} We draw $\eta$ uniformly between $0.8h$ and $1.2h$ once per update and repeat this step the number of times in table~\ref{tab:mc-runs}. The resulting proposal $(q_*,p_*)$ is accepted with Metropolis probability \begin{equation}\label{eqn:mc-acceptance} P_{\rm acc}=\min\!\left\{1,\exp\!\left[\mathcal H_{\rm HMC}(q,p)-\mathcal H_{\rm HMC}(q_*,p_*)\right]\right\}, \end{equation} where $(q,p)$ denotes the initial path and freshly drawn momentum; otherwise we retain the previous path. The HMC step $\eta$ is distinct from the imaginary-time spacing $\epsilon$. Production runs use $\beta=4$ and $\epsilon=0.1$, with sizes and nominal steps $h$ listed in table~\ref{tab:mc-runs}. Measurements follow every update, including rejections. We compute covariances separately at each slice and average over slices at least one unit of imaginary time from either endpoint.

\begin{table}[tbp]
\centering\small
\begin{tabular}{rrrrr}
\toprule
$N$ & Updates, $\lambda=1$ & Updates, $\lambda=100$ & HMC step $h$ & Leapfrog steps\\
\midrule
4  & 60\,000 & 263\,200 & 0.35 & 5\\
6  & 60\,000 & 440\,700 & 0.30 & 6\\
8  & 40\,000 & 56\,600  & 0.26 & 7\\
10 & 30\,000 & 171\,700 & 0.22 & 8\\
12 & ---    & 131\,520 & 0.20 & 8\\
\bottomrule
\end{tabular}
\caption{Total recorded production updates used in the large-$N$ fits, summed over separate Monte Carlo runs at each $N$ and coupling. Newly initialized runs discard respectively $1200,900,600,300,180$ initial updates; these are not included in the table. Acceptance fractions are approximately $0.53$--$0.60$.}
\label{tab:mc-runs}
\end{table}

\subsection{Reference Subtraction and Uncertainties}
To reduce statistical noise, we measure a reference covariance on the same paths. Define the pointwise local energy and reference Hamiltonian by
\begin{equation}\label{eqn:mc-local-energy}
E_{\rm loc}=\Psi_T^{-1}\widetilde H\Psi_T
=U-\frac{|\nabla W|^2-\Delta W}{2N},\qquad
\widetilde H_T=\widetilde H-E_{\rm loc}.
\end{equation}
Here $\nabla$ and $\Delta$ act on the independent traceless components of the single matrix pair $Q$. With a suitable normalizable trial state, $\widetilde H_T$ has $\Psi_T$ as its zero-energy ground state. In its path integral, integrating over all slices except one gives a distribution at that slice proportional to $|\Psi_T|^2$, up to discretization error. The reference average is obtained by reweighting:
\begin{equation}\label{eqn:mc-reweighting}
r[Q]=\exp\!\left[\epsilon\sum_{j=0}^{M}\alpha_j E_{\rm loc}(Q_j)\right],\qquad
\langle O\rangle_{\rm ref}=\frac{\langle rO\rangle_{\rm proj}}{\langle r\rangle_{\rm proj}}.
\end{equation}
Here $O$ is an observable measured at a slice, such as $T_a$ or $T_aT_b$. The subscript $\mathrm{proj}$ denotes an average in the sampled weight $e^{-S_\epsilon}$, and $\mathrm{ref}$ denotes the normalized reference average. We combine separate Monte Carlo runs with the same simulation parameters by adding their recorded sums of $T_a$ and $T_aT_b$. For projected averages, we divide by the total number of retained measurements; for reference averages, we use sums weighted by $r$ and divide by the total weight. To keep the weights from becoming too large for numerical evaluation, we subtract a constant from the exponent in $r$; weights from different runs are rescaled to a common subtraction constant before being combined. We form covariances from these combined means and mean products at each slice, then average over the central measurement window. We fit the difference between projected and reference covariances as
\begin{equation}\label{eqn:mc-extrapolation}
\begin{aligned}
D_{ab}(N)&=\widetilde C^{\rm proj}_{ab}(N)-\widetilde C^{\rm ref}_{ab}(N)
=D_{ab}^{(\infty)}+\frac{a_{ab}}{N^2}+\frac{b_{ab}}{N^4},\\
\widehat C_{ab}(\lambda)&=\lambda^{-(d_a+d_b)/6}
\left(\widetilde C^{\rm SD}_{T,ab}+D_{ab}^{(\infty)}\right).
\end{aligned}
\end{equation}
Here $a_{ab},b_{ab}$ are fit coefficients, $d_a$ is the degree of $T_a$, and tildes indicate rescaled units. The trial covariance $\widetilde C^{\rm SD}_T$ comes from the planar SD calculation, including $B$ where present. Thus the reported estimate retains this external numerical input.

We estimate statistical errors separately for each covariance pair and $N$. A delete-block jackknife divides the ordered data into $n$ contiguous blocks and recomputes the full covariance difference with each block omitted. Denoting these estimates by $D^{(-\ell)}$, their mean and the estimated error are
\begin{equation}\label{eqn:mc-jackknife-error}
\overline D^{(-)}=\frac1n\sum_{\ell=1}^{n}D^{(-\ell)},\qquad
\sigma_{J,n}^{2}=\frac{n-1}{n}\sum_{\ell=1}^{n}\left(D^{(-\ell)}-\overline D^{(-)}\right)^2.
\end{equation}
We retain the larger error from $n=10$ and $n=30$, denoted $\sigma_J$. To account for additional variation across longer segments, also split the data into eight contiguous batches. Let $D_\ell$ be the estimate from batch $\ell$ and $e_\ell$ its ten-block jackknife error. We use
\begin{equation}\label{eqn:mc-batch-error}
\begin{gathered}
\overline D=\frac{\sum_{\ell=1}^{8}D_\ell/e_\ell^2}{\sum_{\ell=1}^{8}1/e_\ell^2},\\
\sigma_D=\sigma_J\max\!\left\{1,\sqrt{\frac17\sum_{\ell=1}^{8}\frac{(D_\ell-\overline D)^2}{e_\ell^2}}\right\}.
\end{gathered}
\end{equation}
The central value remains the estimate from all data; the batch comparison only enlarges its error. The central extrapolation is a quadratic fit in $1/N^2$, using inverse-variance weights $1/\sigma_D^2$ at each $N$. The largest absolute shift to linear fits using all sizes or only $N\geq6$ estimates fit dependence.

At $\lambda=100$, we subtract the covariance difference measured with $\epsilon=0.1$ from that measured with $\epsilon=0.05$, separately at $N=4,6$. Each change has error equal to the square root of the sum of the squared errors of the two independent runs. Their inverse-variance weighted mean is denoted $\delta$; its statistical error $\sigma_\delta$ is the inverse square root of the sum of those weights. Assuming an $N$-independent $O(\epsilon^2)$ bias, the estimated continuum shift and time-step uncertainty are
\begin{equation}\label{eqn:mc-time-error}
s=\frac43\delta,\qquad \sigma_{\rm step}=\frac43\sqrt{\delta^2+\sigma_\delta^2}.
\end{equation}
Halving $\epsilon$ reduces the leading bias by a factor of four, so the measured change is three quarters of the correction from $\epsilon=0.1$ to zero spacing. This gives the factor $4/3$. The plotted centers retain $\epsilon=0.1$. We combine $\sigma_{\rm step}$ with the statistical error of the fitted intercept and the fit-dependence estimate by taking the square root of the sum of their squares, then rescale the error by the same factor as the central value in \eqref{eqn:mc-extrapolation}. The $\lambda=1$ bars contain no additional time-step term. Runs with $\beta=6$ at $N=6$ and changes of the central measurement window provide projection checks, not rigorous bounds on the remaining bias.

\subsection{Normalizability and Trajectory Diagnostics}\label{app:mc-normalizability}
Stable sampling does not establish global normalizability. Figure~\ref{fig:mc-trajectories} shows two recorded midpoint histories and the squared trial wavefunction evaluated as we vary the overall scale of a fixed pair of matrices. At $N=4$, we use
\begin{equation}\label{eqn:mc-ray}
\begin{gathered}
\widetilde X=a\,\operatorname{diag}(1,-1,0,0),\qquad
\widetilde Y_{12}=\widetilde Y_{21}=a,\\
u=\left[\Tr(\widetilde X^4+\widetilde Y^4)\right]^{1/4}=\sqrt2\,|a|,
\end{gathered}
\end{equation}
with all other entries of $\widetilde Y$ zero and $a\geq0$. Increasing $a$ scales both matrices together while keeping their relative entries fixed. The lower panels evaluate the complete function $W$ used in the Monte Carlo runs, including the auxiliary correction where present, through \begin{equation}\label{eqn:mc-wavefunction-scale} \log\frac{|\Psi_T(Q(a))|^2}{|\Psi_T(0)|^2}=-2\bigl[W(Q(a))-W(0)\bigr]. \end{equation} Here $Q(a)$ denotes the pair of matrices in \eqref{eqn:mc-ray}. The coefficient of the highest power of $a$ in $W(Q(a))$ is negative: the power is six at $\lambda=1$ and eight, after including $B$, at $\lambda=100$. This coefficient is also negative for nearby choices of the relative matrix entries, so the norms diverge. Neither positive local fluctuation covariances nor the confining physical Hamiltonian repairs these trial tails in the discretized endpoint weight.

\begin{figure}[tbp]
\centering
\includegraphics[width=\textwidth]{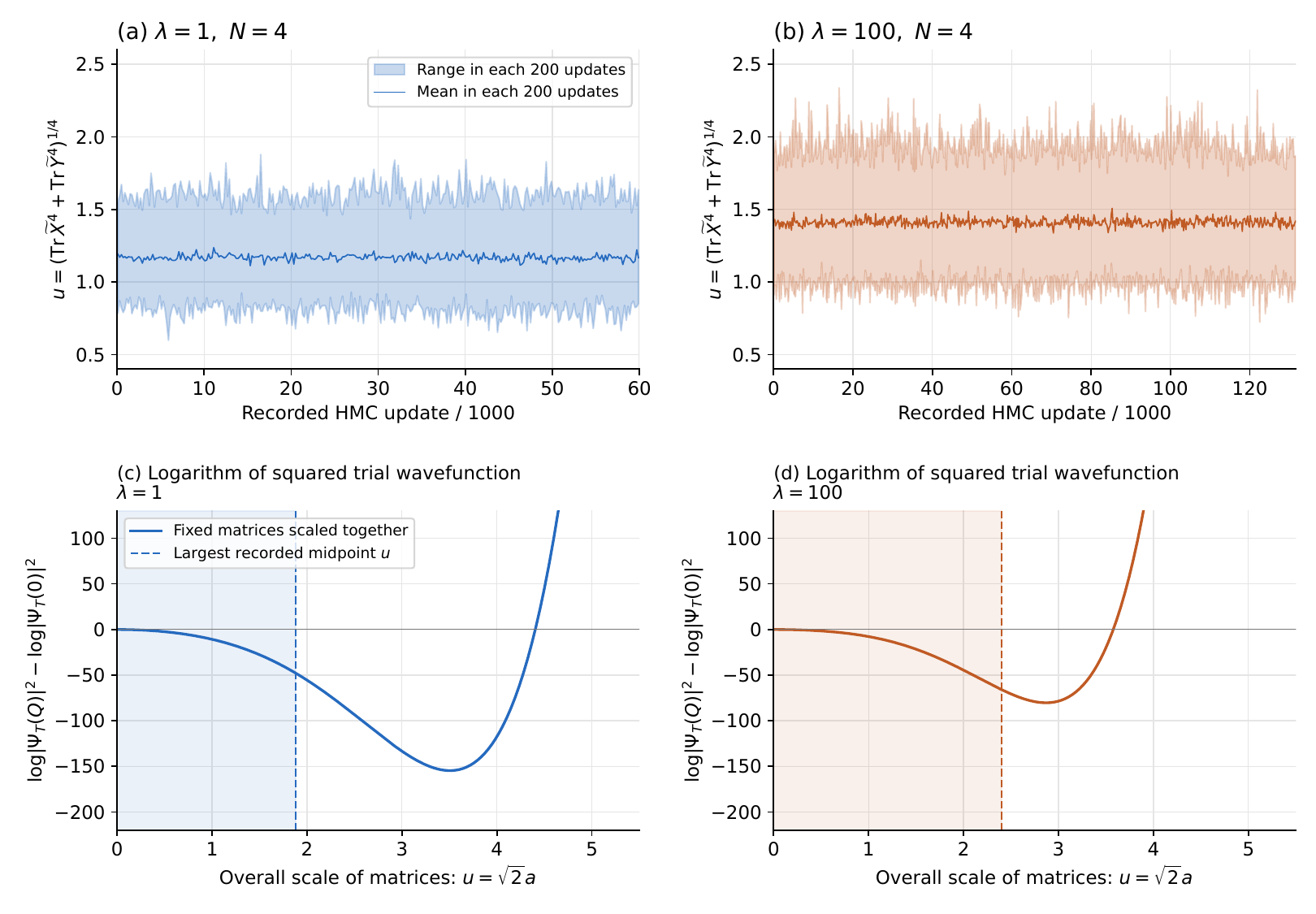}
\caption{\textbf{Top:} midpoint histories at $N=4$, $\beta=4$, $\epsilon=0.1$, with 60\,000 and 131\,600 HMC updates. Lines and envelopes show the mean and full range in groups of 200 updates; the horizontal axis is Monte Carlo update count. \textbf{Bottom:} logarithm of the squared trial wavefunction relative to its value at zero matrices, \eqref{eqn:mc-wavefunction-scale}. As $a$ increases in \eqref{eqn:mc-ray}, both matrices grow with their relative entries fixed; the horizontal coordinate $u$ measures their overall scale. The curves use the Monte Carlo trial-state coefficients, including $B$ at $\lambda=100$. Dashed lines mark the largest midpoint $u$ across the corresponding saved $N=4$ runs, including refinements. These are not escape thresholds: different matrix configurations with the same $u$ can have different wavefunction values, and other ways of varying the matrices could encounter a smaller barrier.}
\label{fig:mc-trajectories}
\end{figure}

The stored midpoint histories show no visible escape. Full endpoint histories were not retained, although some checkpoints contain the final whole path. These records establish finite-time stability in the visited region; they do not bound rare excursions, establish global equilibration, or convert a barrier along this chosen family of matrices into an exponentially small error. In particular, the reference-state identity above is not justified globally by the pointwise equation $\widetilde H_T\Psi_T=0$ alone.

A smooth cutoff $\chi$, equal to one in the central region and vanishing outside a bounded region, would define a normalizable trial state. It also changes the reference Hamiltonian. Where $\chi>0$,
\begin{equation}\label{eqn:mc-cutoff}
\Psi_{T,c}=\chi\Psi_T,\qquad
E_{{\rm loc},c}=E_{\rm loc}
-\frac{\nabla\chi\cdot\nabla\log\Psi_T}{N\chi}
-\frac{\Delta\chi}{2N\chi}.
\end{equation}
Cutoff sensitivity must be tested in both the projected and consistently regulated reference calculations. This remains undone. The quoted bars omit this uncertainty and error in the planar SD reference; the comparisons provide supporting numerical evidence.
\FloatBarrier
\endgroup

\bibliographystyle{JHEP}
\bibliography{refs}

\end{document}